\documentclass{jfm}
\usepackage{graphicx}
\usepackage{nicefrac}
\usepackage{xcolor}
\usepackage{comment}
\usepackage{chemformula}

\shorttitle{Shock ignition and DDT with fluctuations}
\shortauthor{W. Wang, J. McDonald and M. I. Radulescu}

\title{Shock induced ignition and transition to detonation in the presence of mechanically induced non-linear acoustic forcing}

\author{Wentian Wang\aff{1}
 \corresp{\email{Wentian.Wang@uottawa.ca}},
 James McDonald\aff{1},
 \and Matei Ioan Radulescu\aff{1}
  \corresp{\email{matei@uottawa.ca}} 
 }

\affiliation{\aff{1}Department of Mechanical Engineering, University of Ottawa, 161 Louis-Pasteur, Ottawa, K1N 6N5, Canada}

\begin{document}

\maketitle
\begin{abstract}
We address the problem of shock induced ignition and transition to detonation in a reactive medium in the presence of mechanically induced fluctuations by a moving oscillating piston.  For the inert problem prior to ignition, we provide a novel closed form model in Lagrangian coordinates for the generation of the train of compression and expansions, their steepening into a train of N-shock waves and their reflection on the lead shock, as well as the distribution energy dissipation rate in the induction zone.  The model is found in excellent agreement with numerics.  Reactive calculations were performed for hydrogen and ethylene fuels using a novel high-fidelity scheme to solve the reactive Euler equations written in Lagrangian coordinates. Different regimes of ignition and transition to detonation, controlled by the time scale of the forcing and the two time scales of the chemistry: the induction and reaction times. Two novel hot spot cascade mechanisms were identified. The first relies on the coherence between the sequence of hot spot formation set by the piston forcing and forward wave interaction with the lead shock, generalizing the classic runaway in fast flames. The second hot spot cascade is triggered by the feedback between the pressure pulse generated by the first generation hot spot cascade and the shock. For slow forcing, the sensitization is through a modification to the classic run-away process, while the high frequency regime leads to very localized sub-critical hot-spot formation controlled by the cumulative energy dissipation of the first generation shocks at a distance comparable to the shock formation location.   
\end{abstract}

\begin{keywords}
Detonation ignition, deflagration-to-detonation transition, hot spots, spontaneous wave, shock train, Lagrangian reactive gasdynamics 
\end{keywords}
\section{Introduction}
\label{intro}
The process of deflagration to detonation transition (DDT) is central to both astrophysical reactive hydrodynamics, as it controls, for example, Supernovae explosions of white dwarfs, and terrestrial reactive hydrodynamics, as flames, given sufficient development time and appropriate boundary conditions, will eventually accelerate their volumetric burning rate until forming detonations.  Detonations in reactive gases propagate at hypersonic speeds, with Mach numbers ranging from 5 to 8 and overpressures in the range of 10 to 20.  Clearly, the transition phenomenon is to be avoided in the process industries, but can be beneficial in propulsion applications utilizing the tremendous power of detonation waves, such as pulse detonation, oblique detonation and rotating detonation engines. 

The transition from deflagration to detonation is marked by a continuous switch in ignition mechanism.  While in deflagrations, ignition is controlled mainly by diffusion of active species and energy, in detonations, diffusion-less auto-ignition relying on gas compression by waves is the main driving mechanism.  The transition from deflagrations to detonations is a three-dimensional phenomenon, and involves the deformation of the flame surface area by non-homogeneous flow, and even disruption of the flamelet structures themselves, when active turbulent time scales are shorter than those of flames.  Reviews of DDT phenomena can be found in the work of \citet{lee1980mechanism}, \citet{ciccarelli2008flame}, with an entry point in the more modern literature in the more recent work of \citet{poludnenko2019unified}, \citet{oran2020mechanisms}, \citet{saif2017chapman} and \citet{bychkov2012gas} and their co-workers. 

\begin{figure}
\begin{center}
\includegraphics[width=0.3\textwidth]{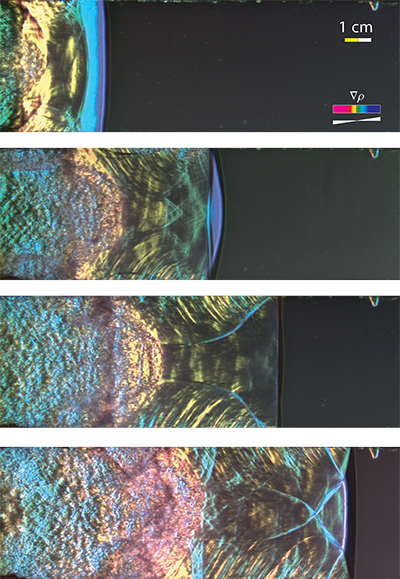}
\caption{The propagation of a Chapman-Jouguet deflagration wave headed by a shock and the subsequent detonation formation in H$_2$-air, $\phi=0.65$, illustrating the pressure waves emitted by the non-planar flame and DDT ahead of the flame in the last frame; adapted from \citet{EderPHD}.}
\label{fig:eder}
\end{center}
\end{figure}

The last stages of DDT generally occur when the volumetric burning rate averaged at some macro-scale associated with the front definition and propagation attains the maximum value permitting steady propagation, denoted by the Chapman-Jouguet condition \citep{eder2001analytical, dorofeev2000effect, chue1993chapman, poludnenko2019unified, saif2017chapman}.  In practice, depending on boundary conditions, this CJ deflagration is invariably headed by a shock wave \citep{rakotoarison2024model}.  An example of the last stages of DDT is shown in Fig.\ \ref{fig:eder}.  At this stage, the flame acts like a fast piston, sustaining the lead shock.  Any acceleration in global burning rate at this stage does not permit quasi-steady propagation and translates into the generation of a train of forward and rear facing shocks.  The DDT process is the series of rapid auto-ignition phenomena that collectively yield a detonation wave.  Experiments and numerical simulations of DDT are usually very difficult to re-construct and rationalize due to this inherent multi-scale and multi-dimensional phenomenon.   This is usually compounded by the fact that hot spot ignition is influenced in a non-trivial way by other neighbouring hot-spots, and their interaction is mainly gas dynamic, i.e., via compression and expansion waves that can be long-ranged. 

DDT onset is also quite difficult to predict in engineering calculations that only assess the global flame dynamics and lead shock strength.  DDT is usually a sub-grid phenomenon to be modelled \citep{dorofeev2000effect, middha2008predicting}.  For example, \citet{meyer1970inadequacy} have shown that the detailed reconstruction of the global dynamics (i.e., space and time filtered) of the lead shock is insufficient to predict the DDT.  More recently, \citet{saif2017chapman} showed that the ignition delays calculated for the mean lead shock speed measured experimentally over-predict the real ignition and DDT time scales observed in the experiments by several orders of magnitude in sensitive mixtures. It is well recognized that this discrepancy is due to fine scale events and hot-spot formations, that can be formed by a variety of mechanisms \citep{saif2017chapman, oran2007origins}.  This calls for a multiple hot-spot model for DDT, which is the main motivation of the present study.

The existing theory of DDT is currently restricted to the formation of single reactive spots in which auto-ignition occurs.  The coupling between the gas-dynamic evolution and the chemical dynamics controlling the ignition delays is now well understood.  It can explain the vast type of behaviour depending on the sensitivity of the induction kinetics to temperature release, initial gradients in fluid state, and time scales of energy deposition.  Reactive front propagation is conditioned by the local ignition time gradients. Much of the work focuses on the coupling of the so-called fast, or diffusionless flames, and the gasdynamics fields, which controls pressure wave amplification. This is a modern extension of the Zel'dovich gradient mechanism that exploits the coherence between the speed of fast flames, controlled by reactivity gradients and acoustic waves.  This coherence extends to shock waves as well, and is sometimes called Shock Wave Amplification by Coherent Energy Release (SWACER) \citep{lee1980mechanism}.  An entry point in the vast modern literature on the subject is the lucid treatment of \citet{sharpe_short_2003}.

The present study aims to extend the previous work on DDT from single hotspots and consider the influence of non-linear acoustic forcing of sufficient high frequency such that multiple hotspots can appear. We thus wish to study the cooperative effect of multiple hot spots on DDT.   We use a 1D approach and consider a non-steady piston, modelling the flame, that generates mechanical disturbances in the medium ahead of it, in the form of compression and expansion waves. 

These waves not only modify the state of the gas ahead of the piston directly, but also interact with the lead shock, generating entropy layers.  The resulting non homogeneous reactive field is then conducive to multiple hot-spot formation and transition to detonation.  The control parameters are the strength of the lead shock, controlled by the mean piston speed and the amplitude and frequency of the mechanical waves, controlled by the piston's speed fluctuations and frequency.  In spite of the model's apparent simplicity, it will be shown numerically and analytically, that different ignition and DDT regimes can be observed, with sometimes profound consequences on the time scales of ignition and DDT and how they differ from calculations without account of perturbations.        

For physical clarity and computational efficiency, the problem is posed in Lagrangian coordinates, such that particle paths are easily tracked and visualized and numerical diffusion usually plaguing this type of numerical calculations can be controlled.  A novel numerical scheme is formulated for reactive gasdynamics in Lagrangian coordinates. We first study the non-steady gasdynamics that arise in a non-reactive medium, treating the problem numerically and theoretically in the weakly non-linear acoustic regime.  The temperature field obtained then serves as leading order solution for investigating and interpreting the reactive dynamics.  The reactive dynamics are determined numerically. We focus on two fuel mixtures, H$_2$-O$_2$ and C$_2$H$_4$-O$_2$, with realistic chemistry to highlight the importance of the different ignition and reaction time scales, as well as the sensitivity of ignition delay to temperature.  

The paper is organized as follows. Section 2 provides the statement of the physical model and governing equations for a reactive diffusionless fluid in Lagrangian coordinates, while a derivation is given in the appendix A.  Section 3 details the formulation of our new a numerical scheme for reactive gasdynamics with multiple species in Lagranaian coordinates and its validation.  Section 4 provides the solution to the inert problem.  Section 5 provides the results of reactive calculations for the two fuels.  Section 6 discusses the various regimes of ignition and DDT in terms of the amplitude and frequency of the forcing. 

\section{Problem definition in Lagrangian coordinates}
\begin{figure}
\begin{center}
\includegraphics[width=0.7\textwidth]{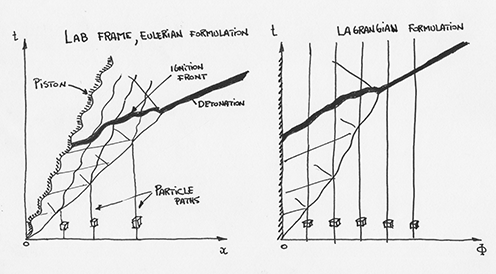}
\caption{Schematic of the problem solved in the Eulerian and Lagrangian frame of reference, in which the particle label $\phi$ serves as independent variable replacing the space variable $x$.}
\label{fig:schematic}
\end{center}
\end{figure}

The one-dimensional problem is illustrated schematically in Fig.\ \ref{fig:schematic} in the laboratory, or Eulerian frame, as well as in the Lagrangian frame of reference following particle paths.  A piston is set in motion at time $t=0$ into a gas at rest and of homogeneous state denoted by a subscript "0".  The piston has a non steady speed given by
\begin{equation}
u_p=u_{p0}+A\sin(2\pi f t)
\end{equation}
where $u_{p0}$, $A$ and $f$ represent the average speed, fluctuation amplitude and fluctuation frequency respectively.  These are held constant. It means the piston motion is modelled as a constant speed motion to which is superimposed a simple harmonic motion (SHM). In current problem, SHM is considered as the fluctuation, i.e., $A \ll u_{p0}$.\\
The piston's impulsive motion generates a main lead shock, followed by a non-homogeneous state affected by the piston fluctuations. These fluctuations are controlled by the compression/expansion waves originating at the piston, interactions of these with the lead shock, generation of entropy waves along particle paths and of course internal and rear boundary and lead shock reflections of these waves.  

The problem can be posed in  Lagrangian coordinates.  A general derivation starting from the more familiar Eulerian formulation is presented in the Appendix A.  The field equations are
\begin{align}
\left(\frac{\partial v}{\partial t}\right)_\phi-\left(\frac{\partial u}{\partial \phi}\right)_t=0 \label{eq:lagrangiancontinity1}\\
\left( \frac{\partial u}{\partial t} \right)_\phi + \left(\frac{\partial p}{\partial \phi}\right)_t=0 \label{eq:lagrangianmomentum1}\\
\left( \frac{\partial e_{tot}}{\partial t} \right)_\phi + \left( \frac{\partial pu }{\partial \phi} \right)_t =0 \label{eq:lagrangianenergy1}\\
\left( \frac{\partial Y_i}{\partial t} \right)_\phi = v \omega_i \label{eq:lagrangianspecie1}
\end{align}
where $v=1/\rho$ is the specific volume, $u$ the particle speed, $p$ the pressure,  $e_{tot}=e+\frac{1}{2}u^2$ is the total energy, $e=\sum\limits_{i=1}^{N}Y_i e_i$ the specific internal energy of the mixture, $Y_i$ the mass fraction of the i$^\text{th}$ component and $\omega_i$ the mass production rate of specie $i$ per unit volume, per unit time, obtained from chemical kinetics.  The independent variables are time $t$ and a mass weighted Lagrangian coordinate $\phi$, defined in terms of the Eulerian space variable $x$ by
\begin{equation}
\phi=\int_{x_p(t)}^x\rho \mathrm{d}x \label{eq:lagrangiantransformation1}
\end{equation}
where $x_p(t)$ is the trajectory of the piston in the laboratory frame.  A line of constant $\phi$ denotes a particle path.  
Partial derivatives with time keeping $\phi$ constant, for example
$\left( \frac{\partial Y_i}{\partial t} \right)_\phi$,
 represent the rate change along a particle path, i.e., a material derivative. 

These equations are supplemented by the equation of state for an ideal gas linking the dependent variables to the mixture temperature $T$ and the prescription of $\omega_i$, the mass production rate of specie $i$ per unit volume per unit time and $e_i$ in terms of the field variables from a thermo-kinetic database.  This formulation is standard and not reproduced here; see for example \citet {kee2005chemically}.   
In the present study, the different chemical components are assumed ideal gases with temperature dependent specific heats entering the prescription of $e_i$.  The \citet{li2004updated} thermochemical database is used for the calculations involving hydrogen and the reduced San Diego thermo-chemical database for the ethylene calculations \citep{varatharajan2002ethylene1,varatharajan2002ethylene2}.  
The use of a reduced mechanism for ethylene is due to the prohibitive computational price of the full San Diego mechanism.  The reduced mechanism provides an accurate prediction as compared to the full mechanism, as we will show below.  A more extensive comparison between the full and reduced mechanisms can be found in the original publications \citep{varatharajan2002ethylene1,varatharajan2002ethylene2}. 
\section{Numerical method}
Both reactive and inert calculations require the numerical solution of the Lagrangian conservation laws given by \eqref{eq:lagrangiancontinity1} to \eqref{eq:lagrangianspecie1}.  We note these are written in conservation form
\begin{equation}
\left(\frac{\partial U}{\partial t}\right)_\phi+\left(\frac{\partial \mathcal{F}}{\partial \phi}\right)_t=Q
\end{equation}
with "conserved" variables $U=\left[v, u, e_{tot}, Y_i \right]^T$, corresponding "fluxes" $  \mathcal{F}=\left[-u, p, pu, 0\right]^T$ and sources $  \mathcal{Q}=\left[0, 0, 0, v \omega_i \right]^T$.  Standard finite volume methods for compressible flow apply, here the volume being a mass element.  While the equivalent modern finite volume methods for hyperbolic system of equations have been demonstrated for Lagrangian coordinates, e.g., \citep{munz1994godunov}, we are not aware of their application to reactive flows.  Our extension of these methods to reactive flows is the well established operator splitting method or Strang splitting method of treating sequentially the inert hydrodynamics and the reactive problems \citep{leveque2002finite}, usually used in solving the Eulerian system of gas-dynamics.  This operator-splitting approach is particularly attractive in the Lagrangian system, as no mass is transferred between the neighbour elements. Each volume contains the same gas through out the entire simulation, which minimizes numerical diffusion errors.  Each element of gas responds solely to compression/expansion from neighbouring ones.   While the inert hydrodynamic solver, detailed next, follows current best practices for high order resolution of gas dynamic discontinuities, the reactive step uses the Cantera package to integrate the resulting ordinary differential equations with their build-in stiff ODE solver \citep{goodwin2017dg}. 
\subsection{Hydrodynamics}
\begin{figure}
\begin{center}
\includegraphics[width=0.6\textwidth]{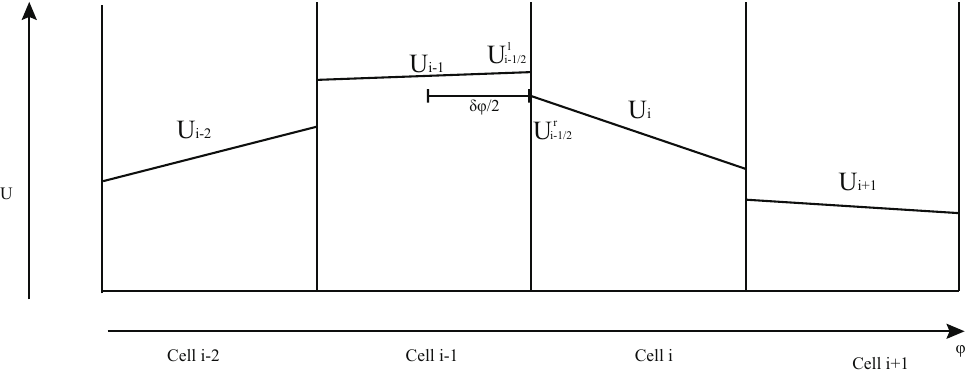}
\caption{At a given time, the solution in neighbouring cell mass elements is represented by piece-wise linear functions used to achieve second order accuracy in the solution of the Riemann problem at cell interfaces for calculating inter-cell fluxes.}
\label{f31}
\end{center}
\end{figure}

In the operator splitting, or Strang splitting approach we use, the contribution of hydrodynamics while keeping a frozen composition is to solve the governing equations  \eqref{eq:lagrangiancontinity1} -\eqref{eq:lagrangianspecie1} without the chemical specie evolution term.  This yields:
\begin{align}
\left(\frac{\partial v}{\partial t}\right)_\phi-\left(\frac{\partial u}{\partial \phi}\right)_t=0 \\
\left( \frac{\partial u}{\partial t} \right)_\phi + \left(\frac{\partial p}{\partial \phi}\right)_t=0 \\
\left( \frac{\partial e_{tot}}{\partial t} \right)_\phi + \left( \frac{\partial pu }{\partial \phi} \right)_t =0 \\
\left( \frac{\partial Y_i}{\partial t} \right)_\phi = 0 
\end{align}
The hydrodynamic step uses a second order HLLE scheme devised for a structured uniform grid in the $\phi$ dimension, as shown schematically in Fig.\ \ref{f31}. The solution vector $U=(v, u, e_{tot}, Y_1, ..., Y_N)$ is stored at the cell centres and the "fluxes" $F=(-u, p,pu, 0, ..., 0)$ are evaluated at the cell interfaces. The solution inside one of these cells is represented by piecewise limited linear functions.  Central differences are used to reconstruct the slopes and van Albada limiter \citep{van1997comparative} has been applied to limit variations in conserved variables. The cell-averaged values of the conserved quantities $U_i$ are updated by computing the flux at the two numerical cell interfaces $\mathcal{F}_{i-1/2}$ and $\mathcal{F}_{i+1/2}$, where $i$ is the index of cells. The inter-cell flux $\mathcal{F}$ is evaluated through the approximation solver of the Riemann problem using the HLLE flux functions with the reconstructed left and right state at the cell boundary described by \cite{einfeldt1991godunov}. The left and right states are interpolated values at the cell of the left and right side of the cell interface. The second order spatial-temporal step is performed as follows.  First a first-order approximation is obtained:
\begin{equation}
\hat{\bar{U}}_{i}^{n+1}=\bar{U}_{i}^{n}+\dfrac{\Delta t}{\Delta x}\left[\mathcal{F}_{i+1/2}^{n}-\mathcal{F}_{i-1/2}^{n}\right]
\end{equation}
This solution $\hat{\bar{U}}_{i}^{n+1}$, together with the initial vector $\bar{U}_{i}^n$ are then used to update the second-order cell state $\widetilde{U}_{i}^{n+1}$ according to
\begin{equation}
\widetilde{U}_{i}^{n+1}=\bar{U}_{i}^{n}+\dfrac{\Delta t}{2\Delta x}\left[\mathcal{F}_{i+1/2}^{n}+\hat{\mathcal{F}}_{i+1/2}^{n}-\mathcal{F}_{i-1/2}^{n}-\hat{\mathcal{F}}_{i-1/2}^{n}\right]
\end{equation}
When the reactions are coupled, $\widetilde{\bar{U}}_{i}^{n+1}$ is further updated after the reaction step. During the hydrodynamic sub step, the ideal gas law is used with an isentropic coefficient, $\gamma$, provided by Cantera at the end of the last chemical sub-step.  The local value of  $\gamma$ is assumed constant during a hydrodynamic step. 

The boundary $\phi=0$ corresponds to the piston face, where the gas speed is that of the piston.  In order to impose a flux into the first cell, the pressure is determined by solving the exact Riemann problem with the velocity of the piston at the cell boundary. Once the velocity and the pressure is evaluated in the cell boundary, the flux can be prescribed.  The boundary condition at the right boundary of the computational domain uses an extrapolation of cubic interpolation; nevertheless, the calculation ends before the lead shock reaches the right boundary such that this local boundary condition does not have any role in the calculation. 
\subsection{Chemical species evolution}
In the operator splitting, or Strang splitting approach we use, the contribution of the change of chemical composition of the gas is to solve the governing equations  \eqref{eq:lagrangiancontinity1} -\eqref{eq:lagrangianspecie1} without the flux terms (derivatives with $\phi$).  This yields:
\begin{align}
\left(\frac{\partial v}{\partial t}\right)_\phi=0 \\
\left( \frac{\partial u}{\partial t} \right)_\phi =0 \\
\left( \frac{\partial e_{tot}}{\partial t} \right)_\phi  =0 \\
\left( \frac{\partial Y_i}{\partial t} \right)_\phi = v \omega_i 
\end{align}
During the reactive step, the specific volume, speed and total energy density (for a fixed mass) are to be kept constant.  This sub step thus naturally lends itself to using Cantera's computational framework already available for this canonical calculation at constant volume and energy. The composition and temperature of each cell is thus evolved as a constant-volume and energy reactor in Cantera during the chemical reaction step.  The ODE's are solved using the SUNDIALS stiff ODE solver used by Cantera over the global time step $\Delta t$ dictated by the hydrodynamics.  

In the Strang splitting approach used, the sequentially taken hydrodynamic and reactive sub-steps thus provide the entire solution vector $(v, u, e_{tot}, Y_1, ..., Y_N)$ at the end of a global time-step.  We use Cantera's build-in solvers to determine the pressure and temperature corresponding to this uniquely defined thermodynamic state by its specific volume, energy and composition.  

\subsection{Numerical verification}
\subsubsection{Verification of the hydrodynamic solver}
\begin{figure}
\begin{center}
\includegraphics[width=0.45\textwidth]{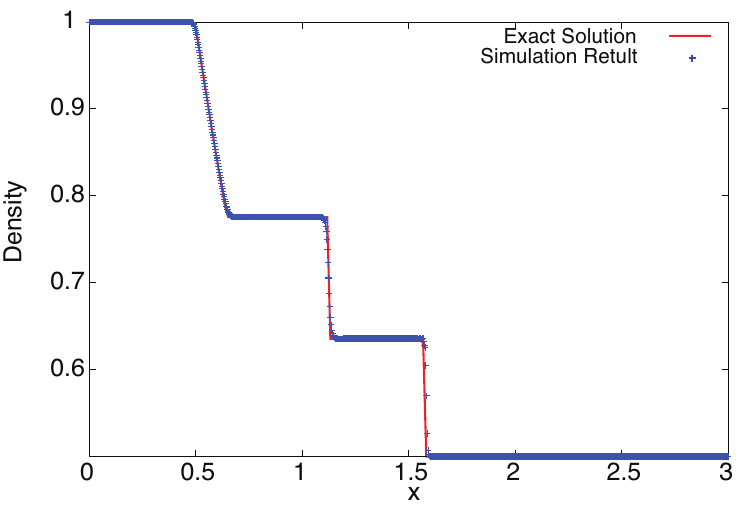}
\includegraphics[width=0.45\textwidth]{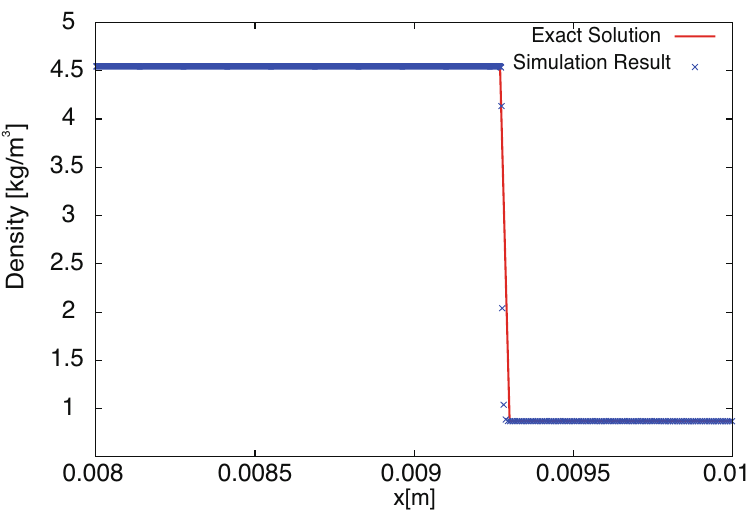}
\caption{Density profile for shock-expansion problem(left); strong shock problem(right)}
\label{f32}
\end{center}
\end{figure}

Two frozen hydrodynamic problems have been chosen to verify the correct implementation of the hydrodynamic solver and confirm the adequacy of the numerical method proposed.  The first problem is a Riemann problem involving a weak shock and expansion fan in a perfect gas.  An initial discontinuity at $x=1$ separates the left state $\left(u_l=0, p_l=2, \rho_l=1\right)$ from the right state $\left(u_r=0, p_r=1, \rho_r=0.5   \right)$ with the isentropic coefficient $\gamma=1.4$ constant for both sides.  Figure \ref{f32} shows the comparison of the calculated density with the exact solution at time $t=0.3$.  It is found that the numerical scheme treats satisfactorily the shock, contact surface and expansion wave typical of the HLLE scheme.  

The second test problem is a strong shock case in a real gas, with frozen chemical composition.  A  piston with a constant speed of 1500 m/s moves into a stoichiometric mixture of hydrogen and oxygen at pressure $p_r=101325$ Pa and temperature $T_r=293$ K.  Figure \ref{f32} shows the comparison of the calculated density with the exact solution at time $t=0.03$ s, where excellent agreement is observed between the computed and the exact solution.  This confirms the reliability of the numerical solver. 

\subsubsection{Verification of the reactive solver coupling to hydrodynamics}
The coupling between the reactive and hydrodynamic solver was tested in the stable detonation wave propagation.  The test is whether the solver can propagate a stable ZND detonation wave.  We tested an over-driven ZND detonation in a stoichiometric mixture of hydrogen and oxygen initially at $1.01\times10^5$ Pa and 300 K.  A piston speed of 2583.6 m/s drives an over-driven detonation wave propagating at 3500 m/s. For reference, the Chapman-Jouguet self-sustained detonation wave speed in this gas is 2839 m/s and the material speed at the CJ state is 1293 m/s.  The driving piston being larger than the CJ material speed gives as expected an over-driven detonation propagating at a speed exceeding the CJ value.

\begin{figure}
\begin{center}
\includegraphics[width=0.7\textwidth]{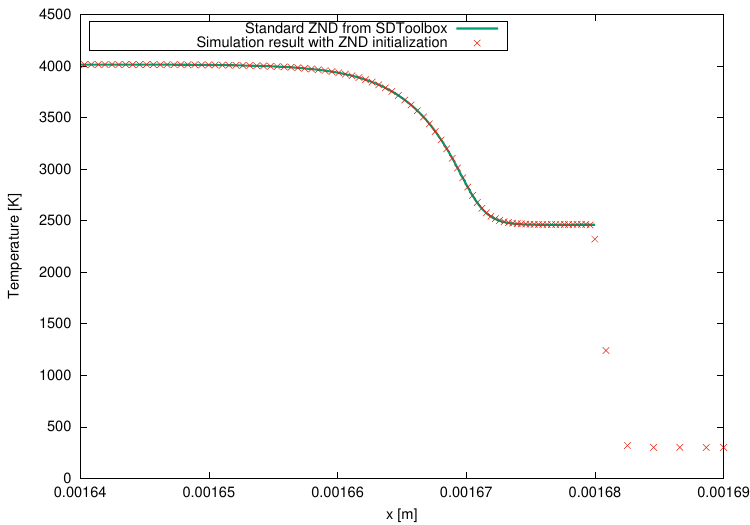}
\caption{Temperature profile for the ZND test}
\label{f33}
\end{center}
\end{figure}

 The calculation was initialized with the over-driven ZND detonation profile calculated separately using the Shock and Detonation Toolbox in Cantera. Figure \ref{f33} shows the evolved wave structure at $t=4.8\times 10^{-7}$ s.  The calculation used a CFL number is 0.7.  Excellent agreement is found between the calculated detonation structure and the expected travelling wave solution provided by the ZND solution, verifying the numerical coupling between the reactive and hydrodynamic solvers. 

\section{The inert problem solution in the induction zone}
\subsection{Overview}
The ignition of the gas induced by the lead shock in the presence of gasdynamic fluctuations is controlled by the temperature and pressure variation in the induction zone, prior to the onset of exothermicity.  It is thus useful to establish the fluctuations induced by the piston first, without account for exothermicity. 

\begin{figure}
\begin{center}
\includegraphics[width=0.7\textwidth]{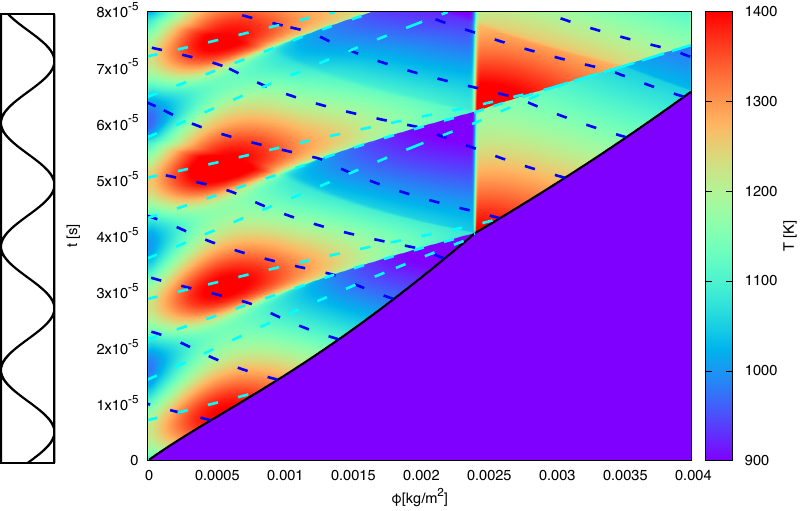}
\caption{Temperature evolution in $\phi-t$ space induced by the impulsive piston motion ($u_0=1626.35$ m/s, $A=0.2u_{p0}$ and $f=45.4$ kHz) into chemically frozen $2H_2+O_2$ initially at $T_0=300$ K and $p_0=5900$ Pa; dark dashed blue lines are C- characteristics while cyan dashed blue lines are C+ characteristics.}
\label{f41}
\end{center}
\end{figure}

A numerical example of the temperature evolution in the induction zone is illustrated in the $\phi-t$ diagram of Fig.\ \ref{f41} by freezing the exothermicity.  The mixture is $2H_2+O_2$ initially at $T_0=300$ K and $p_0=5.90 \times 10^{3}$ Pa, with a mean piston speed $u_{p0}=1.63\times 10^{3}$ m/s, fluctuation amplitude $A=0.2u_0$ and frequency of $f=45.4$ kHz.  This leads to a non-fluctuated post-shock temperature $T_1=1100K$ and post-shock pressure $p_1=1$ Atm.  The reactive solution to this problem is discussed later.   

\begin{figure}
\begin{center}
\includegraphics[width=0.7\textwidth]{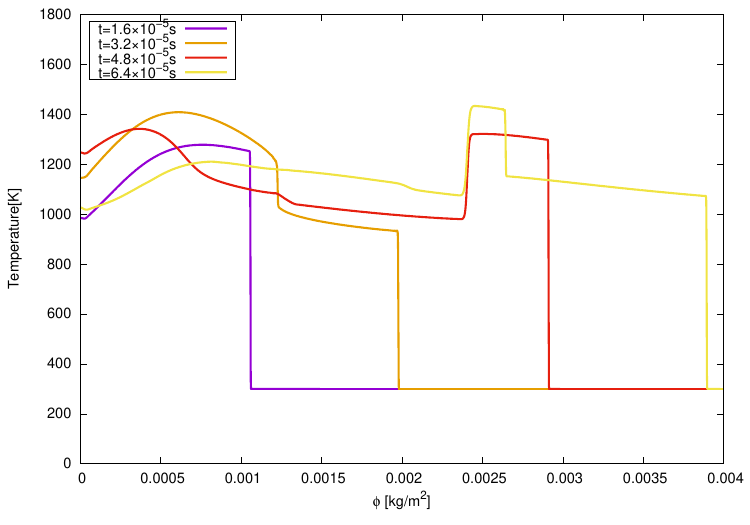}
\caption{Temperature profiles at several times; same conditions as Fig.\ \ref{f41}.}
\label{f42}
\end{center}
\end{figure}
While the impulsive piston motion generates instantly a lead shock, the subsequent compressive parts of the piston motion generate a waveform that steepens to form internal shocks.  These internal shocks can be readily identified as right facing fronts.  This forward facing waveform then reflects on the lead shock, generating reflected disturbances along the C- characteristics and along (vertical) particle paths.  One of these evident entropy waves can be identified as originating at the lead shock when an internal shock wave overtook the lead shock at approximately $t=3.8\times10^{-5}$ s. The particle path evolution can thus be seen to be modulated by the initial temperature obtained at the shock and the subsequent isentropic expansion or compression in regions sufficiently close to the piston before inner shock formation, or further wave by inner shock heating. 

The magnitude and wave shape of the fluctuations can also be observed in Figure \ref{f42}, which shows the temperature distributions at four different times. The last profile corresponds to an instant just prior of the lead shock arrival at the right boundary of the computational domain in this case.  Inner shocks and contact surfaces discussed above can be clearly identified.  Note, however, that it would have been very difficult to rationalize these features without the space-time diagram illustrating the wave dynamics. 

Given the propensity to form inner shocks and their role in shock heating the gas in the induction zone, it is desirable to construct an approximate analytical model for the process to predict the timing of the various events and the temperature amplitude. The problem can be conceptualized as follows. The state generated by an impulsively started steadily moving piston (state 1) behind a constant speed shock can be taken as the leading order constant solution.  Perturbations to this state are brought about by the harmonic piston motion.  Denoting the overpressure of the lead shock as $z_1=(p_1-p_0)/p_0$, the Riemann variable $J^-$ across the shock varies as $O(z_1^3)$ while pressure, density and speed vary as $O(z_1)$ (the acoustic solution) \citep{chandrasekhar1943, whithambook}.  A weakly non-linear description of the state evolution behind the lead shock can thus be sought assuming $J^-=const.$ to order $O(z_1^2)$. This constitutes a "simple wave" solution first suggested by \citet{chandrasekhar1943} and further exploited by \citet{whithambook} in different shock formation problems.  This is the first inert problem to which we seek a simple solution, labelled problem A.  Its weak non-linearity permits to account for the dynamics of inner shock waves. 

The problem B we wish to solve is the interaction of right facing waves described in problem A, with the lead shock, which leads to reflected waves along particle paths and C- characteristics.  This reflection problem is solved in the linear regime and serves to correct the lead shock strength and the interior states provided by problem A. 

We thus seek solutions of the form:
\begin{eqnarray}
u=u_1+u_A+u_B...\\
p=p_1+p_A+p_B...\\
T=T_1+T_A+T_B...
\end{eqnarray}
\subsection{Problem 0: The constant speed shock driven by the steady piston}
The leading order solution is the shock driven by the impulsively started piston at steady speed $u_{p0}$ into quiescent gas at state 0.  This generates a constant state between the piston and the lead shock labelled with subscript 1, which obeys the usual Rankine-Hugoniot shock jump conditions.  

\subsection{Problem A: right facing wave train and inner shock formation}
The piston speed departure from the steady state $u_A(\phi=0)=u_p-u_{p0}=A\sin(2\pi f t)$ generates a train of waves propagating into the gas at state 1. The weakly non-linear waveform generated by an oscillating piston behind the lead shock can be obtained by neglecting $O(z^3)$ variations in entropy and $J^-$ Riemann variable across the lead shock and all internal shocks.  This \textit{simple wave} problem is solved in closed form in Lagrangian coordinates for a perfect gas. We are not aware of its solution elsewhere, although similar arguments were presented by \citet{chandrasekhar1943} and \citet{whithambook} in other problems in Eulerian coordinates, which makes the analysis more complicated.   

As detailed in the Appendix, the characteristic equations for isentropic flow of a perfect gas are 
\begin{equation}
\left(\frac{\partial J^\pm}{\partial t}\right)_\phi \pm \rho c \left( \frac{\partial J^\pm}{\partial \phi}\right)_t=0
\end{equation}
\noindent where the Riemann variables are $J^\pm=2c/(\gamma-1)\pm u$. Since all C- characteristics originate from the upstream state 0, $J^-=J^{-}_1=J^{-}_0$ is constant everywhere to the level of the current approximation and we have immediately the simple wave relation between flow and sound speed perturbations: 
\begin{equation}
c=c_1+(u-u_1)\frac{\gamma-1}{2} \label{eq:cu}
\end{equation}
Given the flow is isentropic in the current approximation, all variables can be related to the sound speed variation
\begin{equation}
\frac{\rho}{\rho_1}=\left(\frac{p}{p_1}\right)^{\frac{1}{\gamma}}=\left(\frac{T}{T_1}\right)^{\frac{1}{\gamma-1}}=\left(\frac{c}{c_1}\right)^{\frac{2}{\gamma-1}} \label{eq:isentropicrelations}
\end{equation}
and using \eqref{eq:cu}, the propagation speed of forward facing characteristics in the $\phi-t$ can be expressed in terms of $u$ only.
\begin{equation}
\frac{d \phi}{dt}=\rho c=\rho_1 c_1 \left(1+\frac{u-u_1}{c_1}\frac{\gamma-1}{2}\right)^{\frac{\gamma+1}{\gamma-1}} \label{eq:cplustrajode}
\end{equation}
Similarly, using \eqref{eq:cu}, the Riemann variable $J^+$ can also be expressed in terms of $u$ only.  Given $J^+$ remains constant along C+ characteristics, it implies that $u$ remains constant along C+ characteristics (and all other variables by virtue of \eqref{eq:cu} and \eqref{eq:isentropicrelations}) and C+ characteristics are straight lines.  These characteristics hence communicate the values of $u$, $c$, $\rho$, $\rho c$,$T$, $p$, etc.\ from the piston face forward. In short, all variables hence satisfy the simple advection equation for $\alpha=u, c, \rho, \rho c,T,p...$,
\begin{equation}
\left( \frac{\partial \alpha}{\partial t} \right)_\phi + \rho_1 c_1 \left(1+\frac{u-u_1}{c_1}\frac{\gamma-1}{2}\right)^{\frac{\gamma+1}{\gamma-1}} \left( \frac{\partial \alpha }{\partial \phi} \right)_t=0 \label{eq:burgerslike}
\end{equation}
The trajectory of any C+ characteristic can be obtained by integrating \eqref{eq:cplustrajode} from the piston $\phi=0$ and reference time $t^*$, yielding 
\begin{equation}
\phi=\rho_1 c_1 \left(1+\frac{u(t^*)-u_1}{c_1}\frac{\gamma-1}{2}\right)^{\frac{\gamma+1}{\gamma-1}} (t-t^*) \label{eq:tstar1}
\end{equation}
Since $u(t^*)$ is the piston speed, this expression provides implicitly the dependence $t^*(\phi, t)$.  Given $t^*$, $u(t^*)$ is known and remains constant along that characteristic.  The other variables are obtained from \eqref{eq:cu} and isentropic relations \eqref{eq:isentropicrelations}. 
Using the harmonic piston speed perturbation, \eqref{eq:tstar1} can be written as 
\begin{equation}
\phi=\rho_1 c_1 \left(1+\frac{A\sin(2\pi f t^*)}{c_1}\frac{\gamma-1}{2}\right)^{\frac{\gamma+1}{\gamma-1}} (t-t^*)
\end{equation}
which can be further simplified for small Mach number harmonic motion or in the Newtonian limit $\gamma \rightarrow 1$ to
\begin{equation}
\phi=\rho_1 c_1 \left(1+  \frac{\gamma+1}{2}\frac{A\sin(2\pi f t^*)}{c_1}\right) (t-t^*)\label{eq:cspeedsimple0}
\end{equation}
The solution is complete but breaks down when characteristics intersect and the solution becomes multi-valued.  When this happens, shocks need to be fitted. The shock formation time and location, as well as the fitted inner shock trajectories can be obtained in closed form. 

\begin{figure}
\begin{center}
\includegraphics[width=0.6\textwidth]{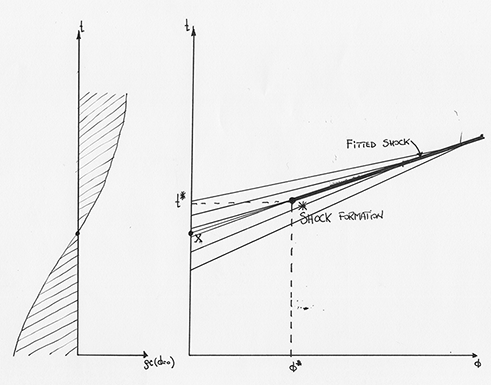}
\caption{Sketch illustrating the shock formation process along the characteristic passing through the point where $\rho c$ had the steepest rate of increase on the piston face (point X); shock forms at point * from characteristics merging.}
\label{fig:shockformation}
\end{center}
\end{figure}

\begin{figure}
\begin{center}
\includegraphics[width=0.45\textwidth]{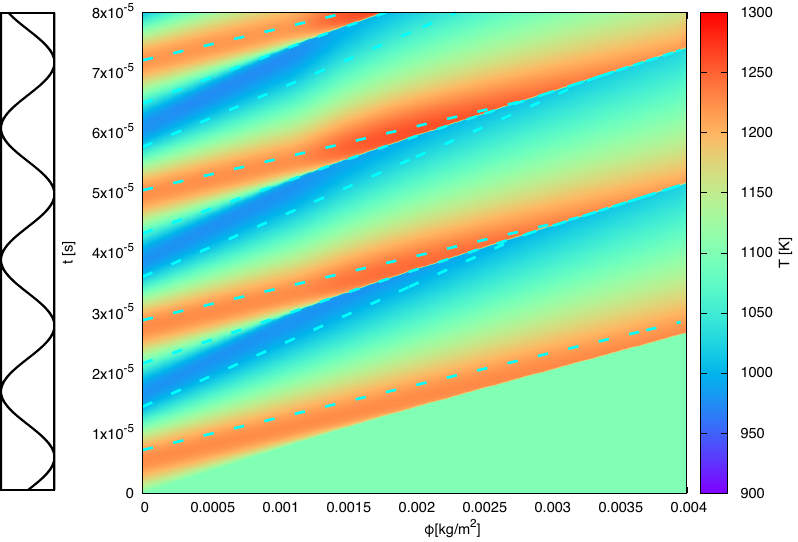}
\includegraphics[width=0.48\textwidth]{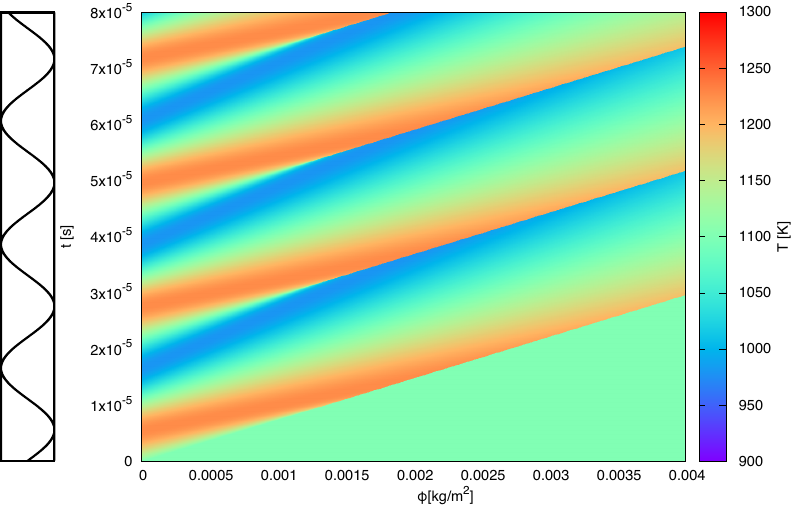}
\caption{Temperature evolution in chemically frozen $2H_2$+$O_2$ initally at $T_1=1100$K generated by a piston speed with fluctuation amplitude $A=325.27$m/s; numerical result (left) and analytical prediction (right).}
\label{f43}
\end{center}
\end{figure}

Figure \ref{fig:shockformation} illustrates the shock formation and the fitted shock trajectory, which approximately bisects the angle formed by the characteristics originating from either side of the shock. Shocks will form along the characteristics where the front was initially steepest at the piston face, i.e. $d(\rho c)/dt$ was initially maximum, since these steepest points on the waveform will be maintained on the same characteristic.  At the piston face, we have 
\begin{equation}
\rho c = \rho_1 c_1 \left(1+  \frac{\gamma+1}{2}\frac{A\sin(2\pi f t^*)}{c_1}\right) \label{eq:cspeedsimple}
\end{equation}
and the steepest points are located at $t=t_{X}=n/f$ with $n=0,1,2, 3....$.  Shocks will thus form along these characteristics, which are given by 
\begin{equation}
(\rho c)_X = \rho_1 c_1 = \frac{\phi}{t-t_X}
\end{equation}
These characteristics will intersect with a neighbouring characteristic originating at point $\phi=0$ and $t=t_X+\delta t$ and having a propagation speed $(\rho c)_X+\frac{d \rho c}{dt} \delta t$ at the shock formation time 
\begin{equation}
t^*=\frac{n}{f} + \frac{1}{(\gamma+1)\pi f A/c_1}
\end{equation}
and shock formation particle label 
\begin{equation}
\phi^*=\frac{\rho_1 c_1}{(\gamma+1)\pi f A/c_1}
\end{equation}
in the limit of $\delta_t\rightarrow0$.  For the example considered, the shock is predicted to form at particle label $\phi^*=1.26\times10^{-3}$kg/m$^2$, in very good agreement with the simulations shown in Fig.\ \ref{f43}. 

The fitted shock trajectory can be approximated to be the continuation of the characteristic along which the shock has first formed.  This approximation is the leading order solution for weak shocks \citep{whithambook}, since the speed of weak shocks is the average of the speed of the characteristics on either side.  It can be shown that the error in shock position location in this case is given by the square of the characteristic speed perturbation, i.e. $\left(\frac{\rho c - \rho_1 c_1}{\rho_1 c_1}   \right)^2$.  This is a higher order effect than the simplification leading to \eqref{eq:cspeedsimple0}, hence can be neglected.  The analytical solution is now complete.  

\begin{figure}
\begin{center}
\includegraphics[width=0.7\textwidth]{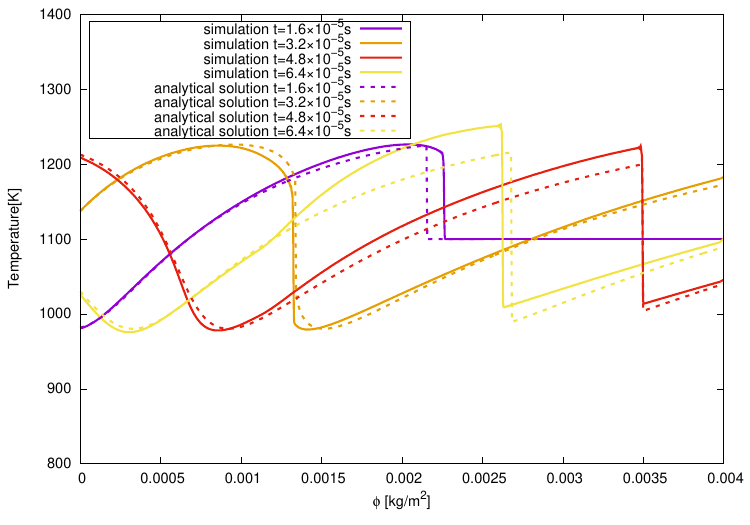}
\caption{Temperature profiles at select times corresponding to the profiles of Fig.\ \ref{f43}.}
\label{f48}
\end{center}
\end{figure}

Figure \ref{f43} shows the prediction of the wave dynamics obtained numerically by the full hydrodynamic numerical solution and its comparison with the analytical prediction developed above. Select profiles are shown in Fig.\ \ref{f48}.  The analytical solution is found in very good agreement with the numerics, in spite of the various simplifications made. The analytical prediction nevertheless somewhat under-predicts the exact numerical solution once the inner shocks are formed. The reason is that the inner shock waves keep dissipating the heat to the local gas during the propagation. Although such heat is small comparing to the fluctuation energy, this mechanism is irreversible and will continuously heat the gas. Before the inner shocks form, this dissipation is negligible and the agreement between numerics and the analytical result is very good. 
\subsection{Decay of the N-waves train}
\begin{figure}
\begin{center}
\includegraphics[width=0.5\textwidth]{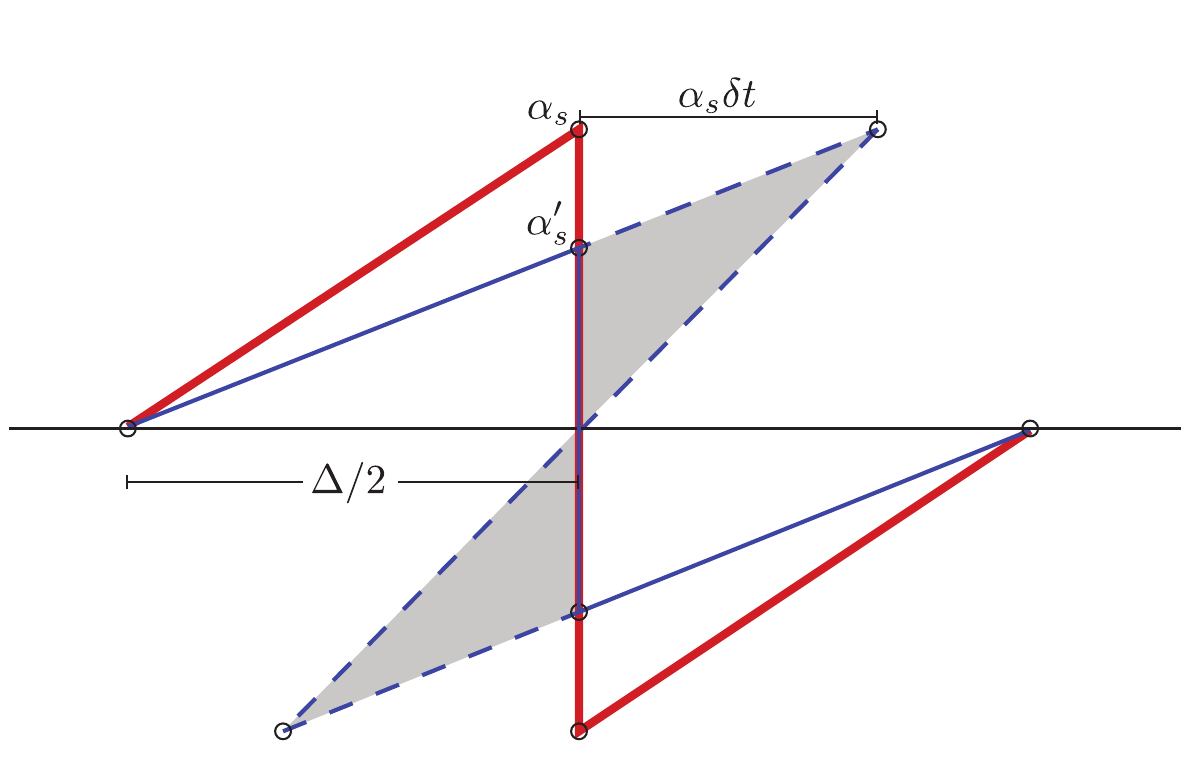}
\caption{
The decay of an N-wave of fixed wavelength $\Delta$; red denotes the wave shape at time $t$ and blue its evolved form after time $\delta t$.  The hatched regions are the equal area lobes justifying the location of the fitted shock at the same position as the initial one.
}
\label{fig:ndecay}
\end{center}
\end{figure}

Once the shocks form, they rapidly asymptote into a train of sequential N-waves that progressively weaken as they travel away from the piston. While their speed is approximately constant, their amplitude is continuously decaying.  It is of interest to establish the decay rate of their amplitude and dependence on forcing frequency, as this serves to understand the reactive solution described below and evaluate their penetration capability.  Consider an N-wave at a sufficiently late time (after travelling a few wavelengths), such that the wave shape can be well approximated by a saw-tooth shape with linear profiles separating the shocks.  The wavelength $\Delta$ is controlled by the forcing frequency, i.e., $\Delta=\rho_1 c_1/f$.  The wave evolution is given by \eqref{eq:burgerslike}; with $\alpha=\rho c$, this is the inviscid Burgers equation.
\begin{equation}
\left( \frac{\partial \alpha}{\partial t} \right)_\phi+\alpha \left(\frac{\partial \alpha}{\partial \phi}\right)_t=0
\end{equation}
In a wave fixed frame moving at speed $\alpha_1=\rho_1 c_1$, the shock is stationary.  After a time $\delta t$, using the method of characteristics, the multi-valued solution requiring shock fitting has displaced the wave shape according to its amplitude $\alpha$; this is shown schematically in Fig.\ \ref{fig:ndecay}. Instead of the multi-valued solution denoted by the wave shape marked by the dotted lines in Fig.\ \ref{fig:ndecay}, the position of the fitted shock coincides with the original one and its new peak amplitude is $\alpha_s'$.  The evolution of the peak amplitude can be easily determined.  In Fig.\ \ref{fig:ndecay}, the slope of the linear profile at time $t+\delta t$ is $\alpha_s'/\left(\Delta/2\right)$.  But since the peak has been displaced by $\alpha_s \delta t$, the slope is also $\alpha_s/\left(\Delta/2+\alpha_s \delta t\right)$.  We thus have the relation
\begin{equation}
\alpha_s'/\left(\Delta/2\right)=\alpha_s/\left(\Delta/2+\alpha_s \delta t\right)
\end{equation}
Writing $\alpha_s'=\alpha_s+\left(d\alpha/dt\right) \delta t$, in the limit of $\delta t \rightarrow 0$, this becomes a differential equation for the wave amplitude
\begin{equation}
\frac{1}{\alpha_s^2}\frac{d\alpha_s}{d t}=-\frac{2}{\Delta}.
\end{equation}
Its solution shows that the N-wave amplitude $\alpha_s$ decays as $\Delta/(2t) =\rho_1c_1/ (2tf)$.
\color{black}

\subsection{Corrections for shock dissipation}
The analytical model described above neglects the energy dissipation of internal shocks.  Our ignition simulations described below show that the cumulative effect of many internal shocks can have a non-negligible effect on ignition, particularly for cases of very high frequency forcing, where an igniting particle undergoes repeated shocking before it ignites.  It is thus of interest to correct the model above and incorporate the energy dissipation of internal shocks.  To this effect, a simple correction scheme is to use the shock strength obtained in the model above, and estimate the irreversible temperature increase via the exact shock jump equations.  
The shock Mach number is evaluated from the jump in particle speed across each shock obtained by the method of characteristics. The resulting temperature increment is the correction that is added to all the gas along that constant-$\phi$ line.  Each successive shock along that line has the same strength and hence the same correction.  For example, gas having been shocked by four internal shocks will have four times that temperature correction. 

\begin{figure}
\begin{center}
\includegraphics[width=0.45\textwidth]{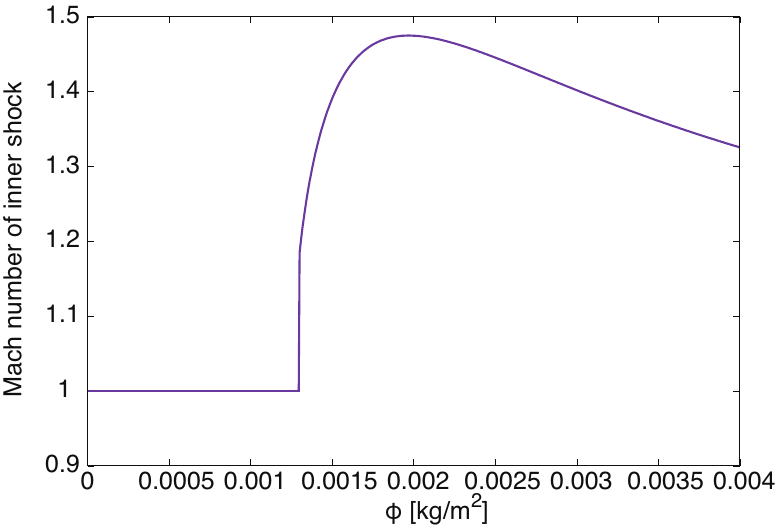}
\includegraphics[width=0.45\textwidth]{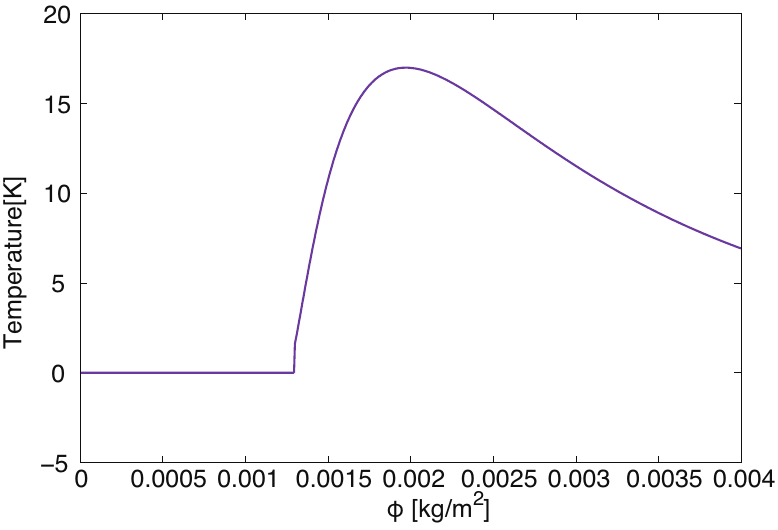}
\caption{Evolution of the Mach number of the internal shock wave (left) and the corresponding irreversible increase across the shock (right).}
\label{f49}
\end{center}
\end{figure}

Figure \ref{f49} left shows the Mach of the inner shocks evaluated as a function of particle label $\phi$.  Figure \ref{f49} right gives the corresponding temperature correction. 
Unity solution in Fig.\ \ref{f49} (left) denotes the region where a shock has not yet formed, in which region there is no shock dissipation and the correction vanishes.

\begin{figure}
\begin{center}
\includegraphics[width=0.7\textwidth]{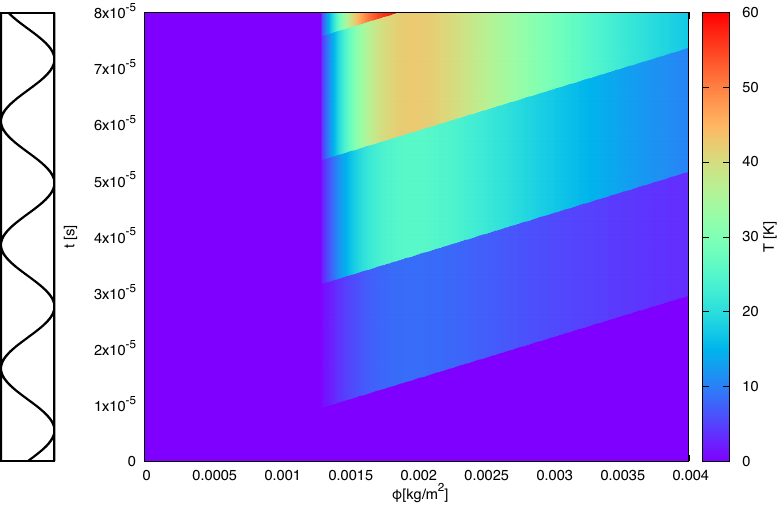}
\caption{The irreversible temperature gain map due to inner shock waves in $2H_2$+$O_2$ mixture at state 1 with $A=325$ m/s.}
\label{f410}
\end{center}
\end{figure}

\begin{figure}
\begin{center}
\includegraphics[width=0.7\textwidth]{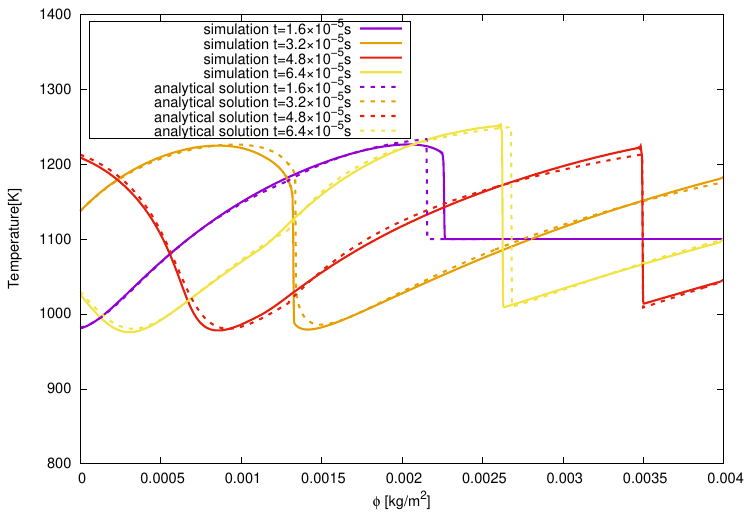}
\caption{Temperature profiles obtained numerically (solid lines) and analytically (broken lines) in $2H_2$+$O_2$ mixture at state 1 with $A=325$ m/s.}
\label{f411}
\end{center}
\end{figure}
The additional temperature correction is shown in Fig.\ \ref{f410}.  The dissipation becomes finite once the inner shocks form at approximately $\phi^*$=0.00126kg/m$^2$ in this case. The increasing of the fluctuation amplitude and the frequency would shorten this distance. To the right of this boundary is the highest region of shock dissipation, which decays as the inner "N" wave structure decay. 

The corrected temperature evolution is shown in Fig.\ \ref{f411}. The temperature amplitude is now in excellent agreement with numerics.  
 
\subsection{Problem B: Reflected waves on the lead shock}
\begin{figure}
\begin{center}
\includegraphics[width=0.7\textwidth]{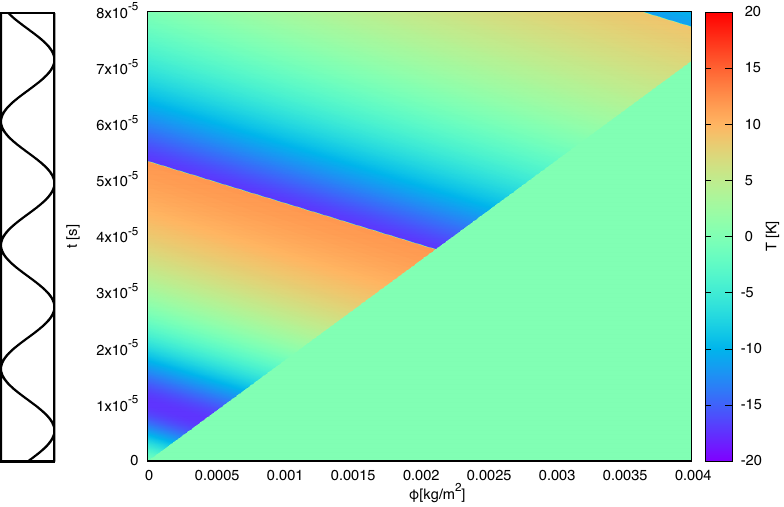}
\caption{Temperature gain contribution due to C- reflected waves due to the reflection of C+ waves interacting with the lead shock in $2H_2$+$O_2$. }
\label{f414}
\end{center}
\end{figure}

\begin{figure}
\begin{center}
\includegraphics[width=0.7\textwidth]{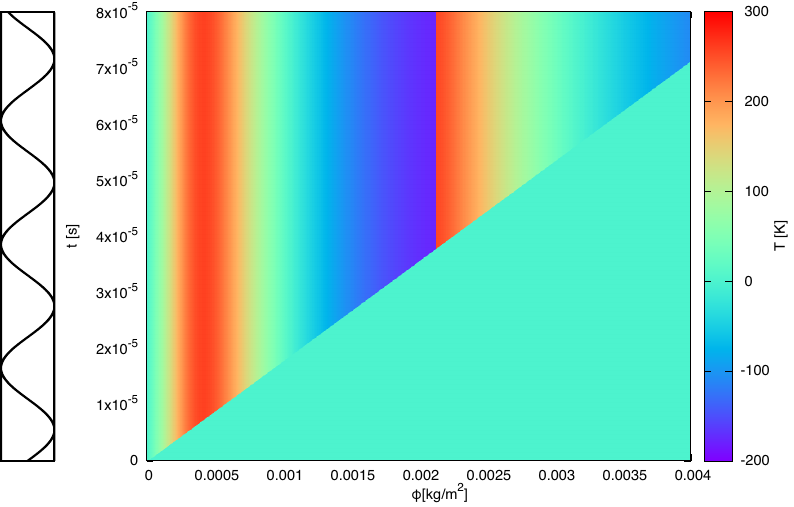}
\caption{Temperature gain contribution due to entropy waves generated at the lead shock from C+ waves reflected at the lead shock in 2$H_2$+O$_2$.}
\label{f415}
\end{center}
\end{figure}

The disturbances generated along the C+ characteristics described above in Problem A finally reflect on the lead shock, generating a wave of opposite nature on the C- characteristics (expansion for an incident compression, and vice-versa) and an temperature (entropy) disturbance on the particle path.  At the level of approximation considered, these reflected disturbances propagate along straight lines $d\phi/dt=-\rho_1 c_1$ for the C- characteristics and $d\phi/dt=0$ for the particle paths by definition and the lead shock path is a straight line.  Appendix \ref{app:shockcatchup} provides the details of this problem for a single disturbance propagating along a C+ characteristic arriving at the lead shock. Note that the reflected acoustic disturbances are much weaker than the entropy waves, hence re-reflections between these waves with other incident waves are not treated. The treatment provided in the Appendix for a single wave can be generalized for the continuum of waves arriving at the lead shock.  Say the lead shock strength at location $\phi$ is known. The C+ characteristic arriving at location $\phi+\delta\phi$ will change the shock strength and provide temperature perturbations along the C- and C0 waves.  These perturbations are then added along these respective lines for the entire domain. Marching along the shock, the arrival of the next C+ characteristic provides the next perturbation and reflected perturbations along C- and C0 lines.  We marched along the shock at regular intervals in order to obtain the entire interior solution.  The interior solution is then interpolated on the grid with the desired resolution. The entire solution for reflected waves can thus be obtained by the method of characteristics.  The temperature perturbations due to the reflected acoustic disturbances are shown in Fig.\ \ref{f414} while the temperature perturbations in Fig.\ \ref{f415} for the chemically frozen 2H$_2$+O$_2$ mixture with post shock temperature of $T_1=1100$ K and fluctuation frequency of $f=45.35$ kHz. 

\begin{figure}
\begin{center}
\includegraphics[width=0.7\textwidth]{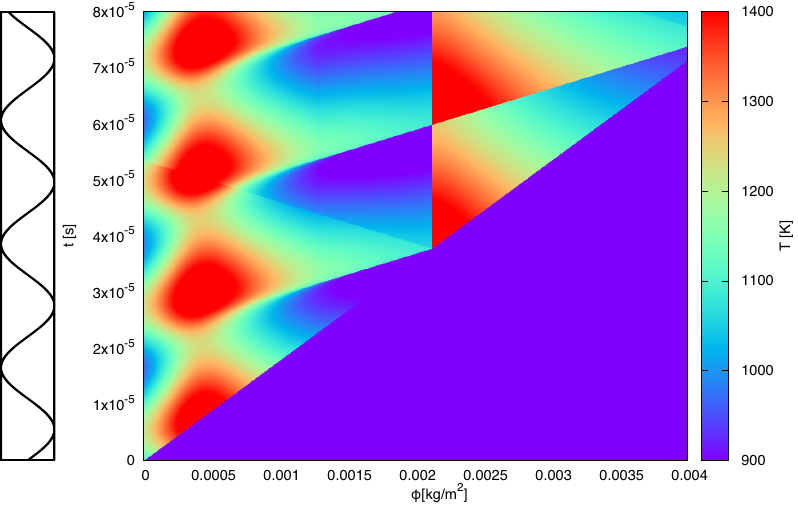}
\caption{Analytically reconstructed temperature field for 2$H_2$+O$_2$ at conditions of Table \ref{tab:mixprops} and $A=0.2 u_{p0}$ and $f=45.4$ kHz without chemical reaction.}
\label{f412}
\end{center}
\end{figure}

These two solutions can be linearly combined along with the solution of problem A to obtain the desired solution for temperature evolution in the induction zone.  The approximate solution illustrated in Fig.\ \ref{f412} is found in general good agreement with the full numerical solution of Fig.\ \ref{f41}.  The timing and amplitude of the various events are well-reproduced.  Note however that the central part of the hot spots are predicted to be approximately 5$\%$ hotter than the full simulation results; likewise, cooler portions are under-predicted by the approximate model by a similar amount. 

\begin{figure}
\begin{center}
\includegraphics[width=0.485\textwidth]{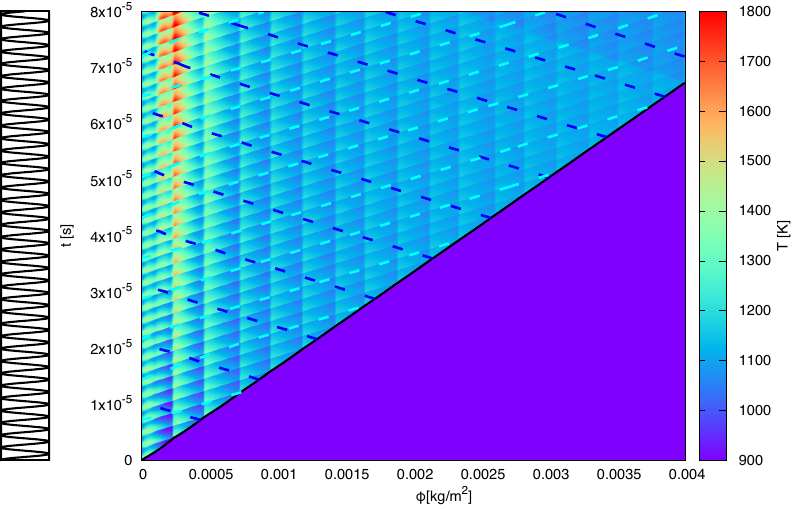}
\includegraphics[width=0.45\textwidth]{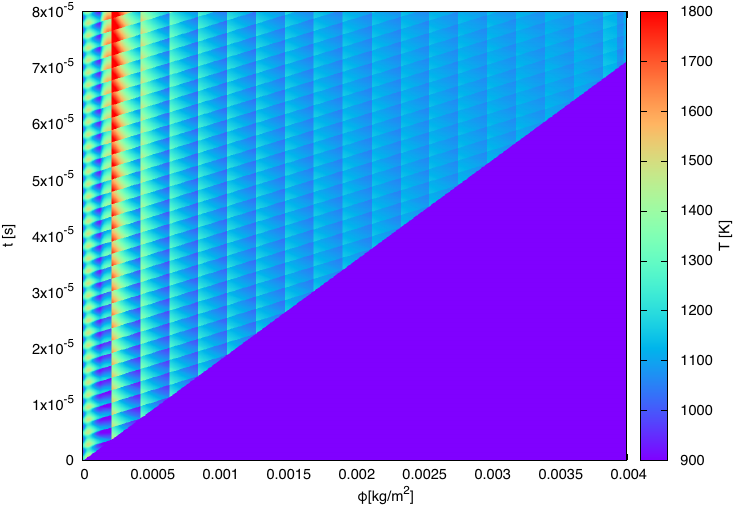}
\caption{Temperature field obtained numerically (left) and analytically (right) for 2$H_2$+O$_2$ at conditions of Table \ref{tab:mixprops}, $A=0.2 u_{p0}$ and $f=454$ kHz without chemical reaction.}
\label{f416}
\end{center}
\end{figure}

The gasdynamic model formulated revealed the most important effects to consider in the interpretation of the reactive problem, namely the propensity to form inner shocks at a finite distance from the piston.  These inner shocks generate entropy and the cumulative effect may generate the strongest hotspots.  Second in order of importance are the reflections of these shocks on the lead shock, generating temperature perturbations by strengthening the lead shock.  Note that the inner shocks decay as N-waves as $1/t$.  In problems with very high frequency, the resulting temperature increase near the piston dominates the temperature increase away from the piston.  An example is shown in Fig. \ref{f416} obtained numerically and using the approximate model developed. In this case, the fluctuation frequency is increased to 10 times the value as the case shown in Fig. \ref{f41}. The strongest hotspot is dominated by the inner shock dissipation. Overall, the analytical results predict very well the location and temperature of the hot spot.

\section{The reactive problem solution}

\subsection{Fuels and their characteristic ignition and reaction times}
\begin{figure}
\begin{center}
\includegraphics[width=0.45\textwidth]{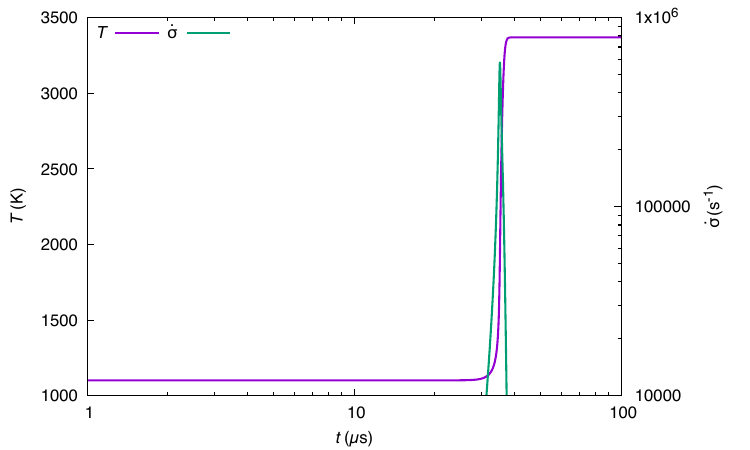}
\includegraphics[width=0.45\textwidth]{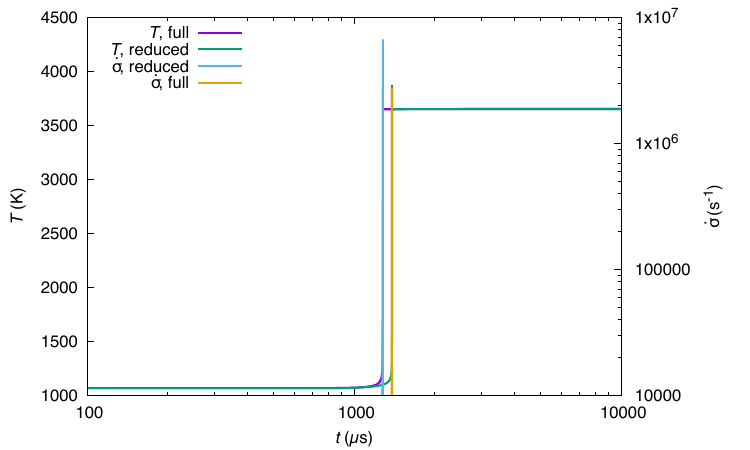}
\caption{
Temperature and thermicity evolution at constant volume for 2$H_2$+O$_2$ (left) and C$_2$H$_4$+3O$_2$ (right) taking state 1 (see Table \ref{tab:mixprops}) as initial condition; "full" and "reduced" profiles denote solutions obtained with the full San Diego mechanism and the reduced mechanism used in this study.}
\label{f51}
\end{center}
\end{figure}

Shock-induced ignition and transition to detonation was studied in two reactive mixtures spanning the different behaviour of ignition observed in practice, namely 2$H_2$+O$_2$ and C$_2$H$_4$+3O$_2$ at temperatures of approximately 1100K, which is the lower end temperatures permitting sufficiently rapid auto-ignition and transition to detonation in practice.  For the 2$H_2$+O$_2$ mixture, a post-shock temperature of $T_1=1100$K and post-shock pressure of $p_1=1$atm are obtained by a piston with mean speed 1626 m/s moving into a gas at an initial temperature of $T_0=300$K and initial pressure of $p_0=6$kPa.  This choice was also motivated by the abundance of experiments on reference ignition data, e.g., induction delay, for hydrogen-oxygen at 1 atm, as well as addressing the final stages of DDT where sufficiently strong shocks are generated in this range of temperatures. Moreover, approximately 1100 K is the lower limit of high-temperature ignition for hydrogen-oxygen ignition at 1 atm \citep{meyer1971shock}.  The test conditions for C$_2$H$_4$+3O$_2$ were $T_0=300$K and $p_0=6.2$ kPa, with a mean piston of $u_{p0}=1259$m/s, which brings the post shock conditions to those studied by \cite{SaifThesis} in deflagration to detonation transition experiments.  Table \ref{tab:mixprops} lists the relevant mixture properties at these conditions. We first characterized the relevant time scales in the two mixtures in constant volume calculations; the piston forcing characteristics will be referenced to the relevant chemical times.  We define the ignition delay time $t_i$ at constant volume as the time elapsed until maximum thermicity. For a mixture of ideal gases, the thermicity reduces to 
\citep{Fickett&Davis1979, Williams1985, KaoShepherd2008}:  
\begin{equation}
\dot{\sigma}= \sum\limits_{i=1}^{N}\left(\frac{\bar{W}}{W_i}-\frac{h_i}{c_pT}\right) \frac{D Y_i}{D t}
  \label{eq:thermicityidealmix}
\end{equation}
\noindent where $W_i$ is the molecular weight of the i$^\text{th}$ component, $\bar{W}$ is the mean molecular weight of the mixture, $h_i$ is the specific enthalpy of the i$^\text{th}$ specie and $c_p$ is the mixture frozen specific heat and $Y_i$ is the mass fraction of the i$^\text{th}$ specie in the mixture of $N$ total number of species.  The other relevant chemical time is the exothermic reaction time $t_r$, during which energy release affects the dynamics of the gas. It is defined as the inverse of the maximum thermicity. These two time scales are reported in Table \ref{tab:mixprops}. The ethylene mixture is characterized by a much larger ratio of these two time scales $\Lambda=t_i/t_r$.  This value lies at the lower range of most hydrocarbons, which partly motivates our selection of this mixture. Very large values are computationally prohibitive, since the reaction time and length scales require sufficient resolution.  Figure \ref{f51} shows the evolution of temperature and thermicity for the two fuels.  
For ethylene, we have also calculated the ignition process using the full mechanism.  The reduced mechanism predicts well the ignition delay, with a minor under-prediction.  Both models predict very short reaction times $t_r$, the reduced model over-predicting it by a factor of 2.   In the reactive calculations presented below, we ensured that the reaction zone thickness was sufficiently well resolved with at least 10 points.  The grid spacing used $\Delta_\phi$ was $2\times10^{-6}$kg/m$^2$ for the hydrogen mixture and $2.8\times10^{-6}$kg/m$^2$ for the ethylene mixture. A detailed grid sensitivity and convergence study is reported in Appendix \ref{app:convergence}.

For reference, we also provide the calculations of the hypothetical CJ detonation and the corresponding Von Neumann state.  These values are tabulated in Table \ref{tab:mixprops}.  Of interest are the values of material velocity at the CJ state.  Since in both cases the piston speed is larger than the material speed at the CJ state, this means that the detonation waves established by the piston speeds selected are overdriven.  

\begin{table}
 \begin{center}
 \caption{
Relevant thermo-chemical properties of the two reacting mixtures.
 \label{tab:mixprops}}
  \begin{tabular}{lcc}
  \hline
    &  2H$_2$+O$_2$ &    C$_2$H$_4$+3O$_2$\\
    \hline
\textbf{initial state (0)}\\
    
    temperature $T_0$ & 300 K & 300 K \\
    pressure $p_0$  & 5.9$\times 10^3$ Pa & 6.2$\times 10^3$ Pa \\
    \hline  
    shock speed $D_s$  & 2.06 $\times 10^3$m/s & 1.47 $\times 10^3$ m/s\\
    piston speed $u_{p0}$  & 1.63 $\times 10^3$ m/s & 1.26 $\times 10^3$ m/s \\
    \hline
    \textbf{post shock state driven by piston (1)}\\     
    temperature $T_1$ & 1100 K & 1069 K \\
    pressure $p_1$& 1.01$\times 10^5$ Pa & 1.5$\times 10^5$ Pa \\
    density $\rho_1$  &0.133 kg/m$^3$ & 0.522 kg/m$^3$ \\
    sound speed $c_1$  &1.01$\times 10^3$ m/s &5.86$\times 10^2$ m/s \\
    non-dimensional activation energy $T_a/T_1$ & 8.7 & 22 \\
    ignition delay $t_{i0}$ & 3.56$\times 10^{-5}$ s &1.39$\times 10^{-3}$ s\\
    reaction time $t_{r}$ &1.72$\times 10^{-6}$ s & 3.61$\times 10^{-7}$ s\\
    mass weighted reaction thickness\\
     $\delta_{\phi,r}=\rho_1\left(D_s-u_p) t_r \right)$ &2.09$\times 10^{-3}$ kg/m$^2$ & 1.57$\times 10^{-1}$ kg/m$^2$ \\
    \hline
    \textbf{Von Neumann state of CJ detonation}\\     
    shock speed $D_{CJ}$  & 2.69 $\times 10^3$ m/s & 2.24 $\times 10^3$ m/s\\
    material speed $u_{p,VN}$  & 2.19 $\times 10^3$ m/s & 1.99 $\times 10^3$ m/s \\
    temperature $T_{VN}$ & 1616 K & 1874 K \\
    pressure $p_{VN}$& 1.71 $\times 10^5$ Pa & 3.51$\times 10^5$ Pa \\   
       \hline
    \textbf{post detonation state of CJ detonation}\\     
    material speed $u_{p,CJ}$  & 1.23 $\times 10^3$ m/s & 1.04 $\times 10^3$ m/s \\
    temperature $T_{CJ}$ & 3184 K & 3391 K \\
    pressure $p_{CJ}$& 9.84 $\times 10^4$ Pa & 1.85$\times 10^5$ Pa \\   
  \end{tabular}
 \end{center}
\end{table}

\begin{figure}
\begin{center}
\includegraphics[width=0.45\textwidth]{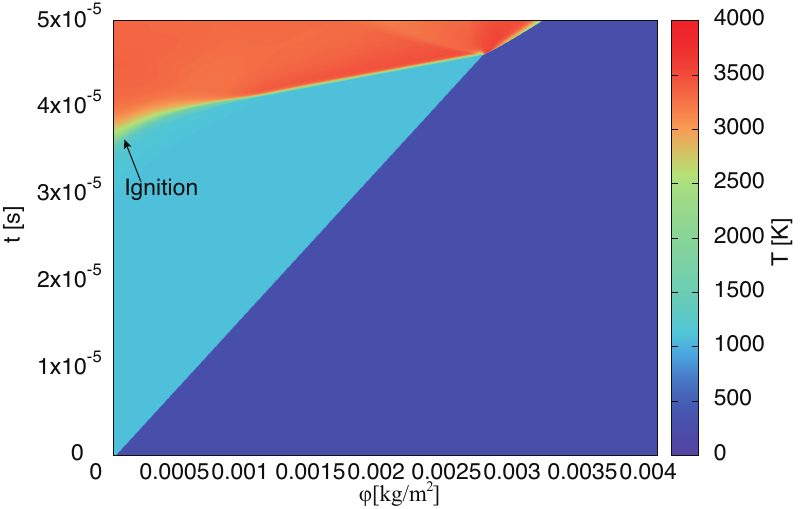}
\includegraphics[width=0.45\textwidth]{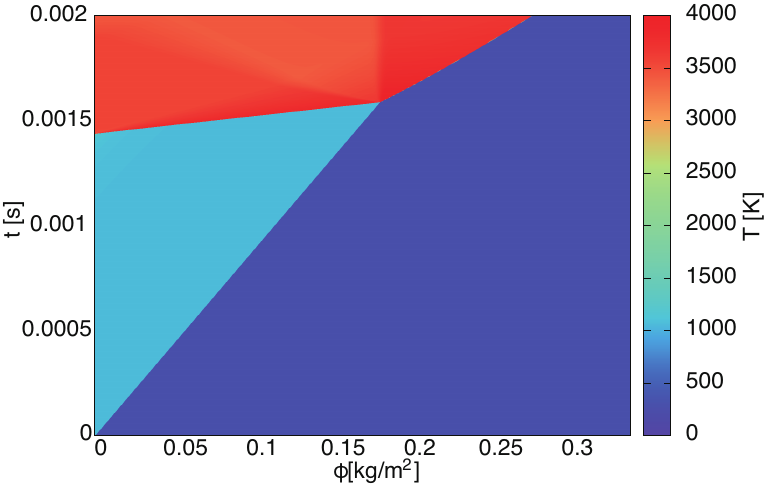}
\caption{Shock induced ignition in 2$H_2$+O$_2$ (left) and C$_2$H$_4$+3O$_2$ (right) without perturbations; the initial conditions in the pre-shock region are zero velocity and state 0, with piston velocities given in Table \ref{tab:mixprops}; the post shock state is state 1.}
\label{f51.2}
\end{center}
\end{figure}
\subsection{Ignition and transition to detonation without fluctuations}
The dynamics of shock induced ignition were first studied in the absence of fluctuations.  Figure \ref{f51.2} shows the evolution of the temperature field.  For the hydrogen mixture, the first ignition along the piston path occurs approximately at the constant volume ignition time, since prior to this event, the lack of thermal evolution keeps the gas shocked by the lead shock at the constant post shock state.  The subsequent rapid acceleration of the reaction zone and formation of an internal shock on the time scale of energy deposition $t_r$ is characteristic of such dynamics; these internal dynamics are compatible with the model of \citet{sharpe2002shock}.  The rapid internal acceleration of the fast flame leads to the development of an internal Chapman-Jouguet detonation wave followed by an interior Taylor wave
 \citep{Fickett&Davis1979}. The arrival of the internal detonation wave at the lead shock transmits an overdriven detonation wave in the quiescent gas, which eventually decays towards the self-sustained overdriven solution supported by the piston speed.

 The transition sequence for the ethylene mixture follows a similar sequence, but the rapid initial transition occurs faster, on time scales of $t_r$, which are much shorter than the ignition delay time scale in this case. 

\subsection{Slow forcing}
\begin{figure}
\begin{center}
\includegraphics[width=0.7\textwidth]{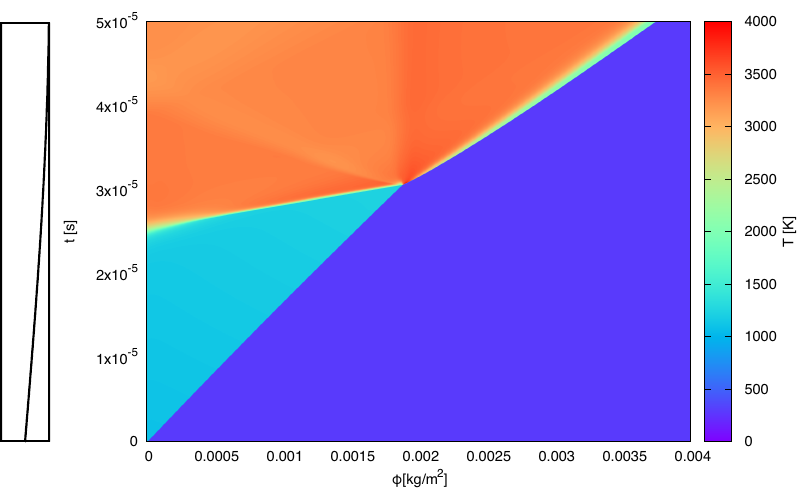}
\caption{Shock induced ignition with 2$H_2$+O$_2$, $f=4.535kHz$, $A=0.2u_p$}
\label{f52}
\end{center}
\end{figure}

When the forcing period is longer than the induction delay, the sequence of events leading to auto-ignition resemble qualitatively the ignition sequence without fluctuations.  An example is shown in Fig. \ref{f52} for the hydrogen system.  The slow compression of the gas in the induction zone shortens the ignition delay $2.5\times 10^{-5}s$.  One could modify the analysis of \citet{sharpe2002shock} to account for the non-uniform conditions in a straightforward, albeit algebraically complex manner. Note the phase of the fluctuation is likely to be very important here.  Initial cooling of the induction zone gas would have the opposite effect to delay the ignition.  The effect of the phase of the fluctuation, while interesting in its own right, is left for future study.

\begin{figure}
\begin{center}
\includegraphics[width=0.7\textwidth]{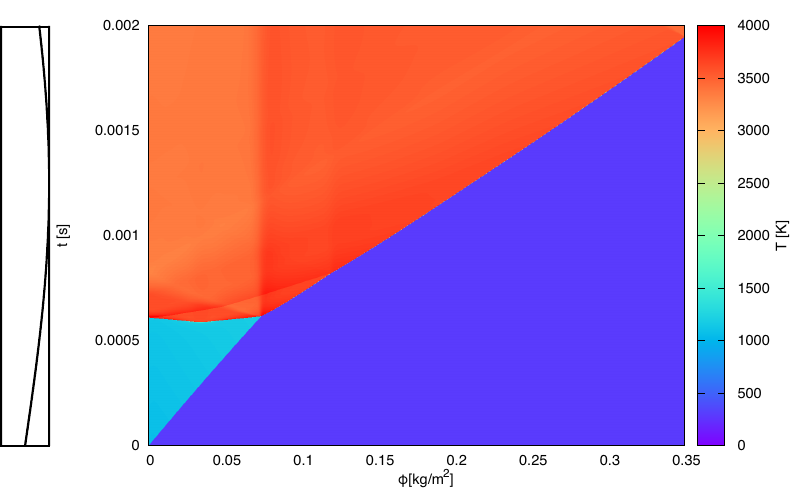}
\caption{Shock induced ignition in C$_2$H$_4$+3O$_2$ with $f=0.2kHz$, $A=0.2u_{p0}$.}
\label{f53}
\end{center}
\end{figure}

When the period of the fluctuation becomes comparable with the induction delay, it modifies the induction time gradient more significantly and can give rise to first ignition away from the piston face.  An example is shown in Fig.\ \ref{f53} for the ethylene mixture.  The gradient of ignition delay set up by the lead shock and modulated by the piston non-steadiness can be flatter than that provided by a steady shock.  Its phase velocity becomes closer to the acoustic signals $\pm \rho_1 c_1$, leading to more prompt acceleration by the Zel'dovich gradient mechanism \citep{sharpe_short_2003} in both directions. While not shown, the slope of the C+ and C- characteristics in the induction zone is $\pm \rho_1 c_1$, which, by virtue of the values of Table 1 for the shock speed, are 2.5 times steeper than the lead shock.  It can be speculated that the rapid formation of inner detonations are due to these more favourable gradients.

\subsection{Forcing on induction time scales}
\begin{figure}
\begin{center}
\includegraphics[width=0.7\textwidth]{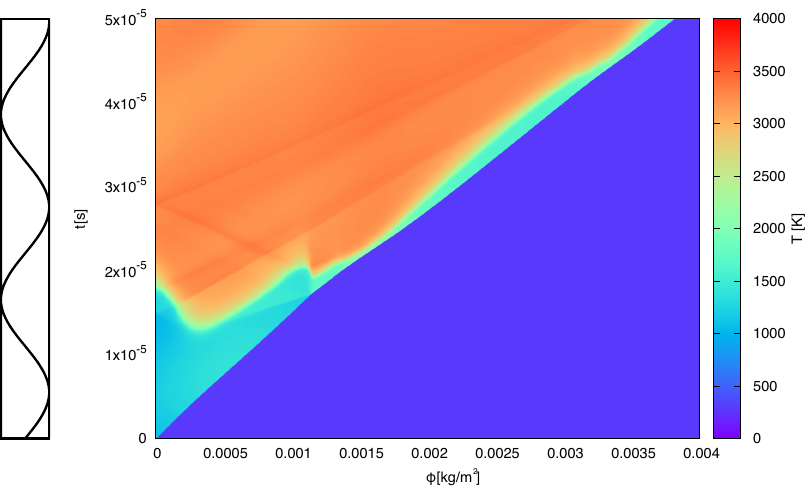}
\caption{Shock induced ignition in $2H_2+O_2$ with $f=45.4$ kHz and $A=0.2 u_{p0}$.}
\label{f541}
\end{center}
\end{figure}
When the forcing frequency is further increased, the non-linearity results in stronger lead shocks by the inert mechanism discussed above, but the phase velocity of the fast flames are now larger than the acoustic speeds, resulting in de-coherence of the Zel'dovich mechanism.  Figure \ref{f541} shows a striking example of this situation.  The first ignition begins at $t=1.3\times 10^{-5}s$. The spontaneous wave from the first ignition generates a forward and backward shock, which are out of phase with the fast flame.  The forward facing shock catches up to the lead shock and generates the second hotspot at $\phi=1.2\times 10^{-3}kg/m^2$. The fast flame evolving from the second hotspot is now in phase with the acoustics, resulting in detonation formation.  After this 1D detonation structure forming, the reaction zone length oscillates almost in phase with the fluctuation, but the reaction never decouples from the lead shock. Some weak shocks can be observed in the reacted gas with some complex forward and backward patterns. They are formed by interaction between the previous inner shock and the boundary (piston or lead shock wave). Since these shock waves are not in the reaction zone, they do not influence the detonation formation process.
\begin{figure}
\begin{center}
\includegraphics[width=0.7\textwidth]{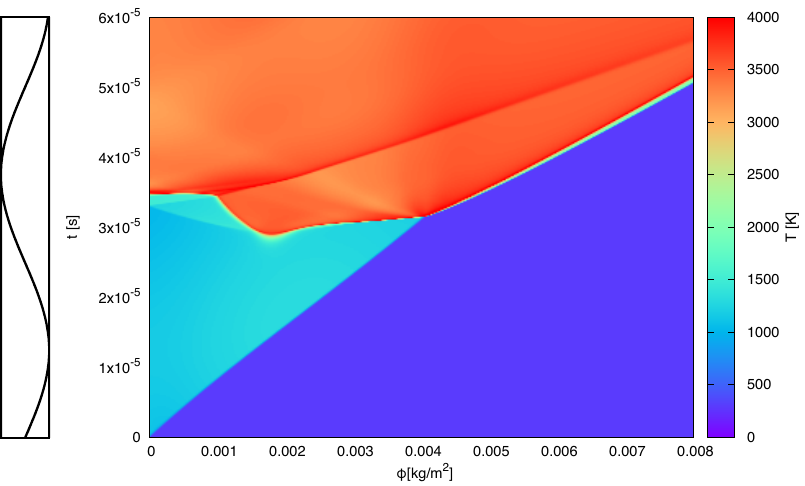}
\caption{Shock induced ignition in C$_2$H$_4$+3O$_2$ with $f=20$ kHz and $A=0.2 u_{p0}$.}
\label{f55}
\end{center}
\end{figure}

A similar hotspot ignition mechanism was also observed in the ethylene-oxygen system.   The first ignition starts on the first hotspot at $t=2.80\times 10^{-5}s$. Different from the hydrogen-oxygen ignition, the spontaneous wave generates two stronger shock waves and the forward propagating shock wave keeps in phase with the reaction front until it reaches the lead shock. The detonation forms prior to the reaction zone reaching the second hotspot. The backward propagating shock wave decouples from the reaction front and reflects on the piston prior to the reaction front.

\begin{figure}
\begin{center}
\includegraphics[width=0.7\textwidth]{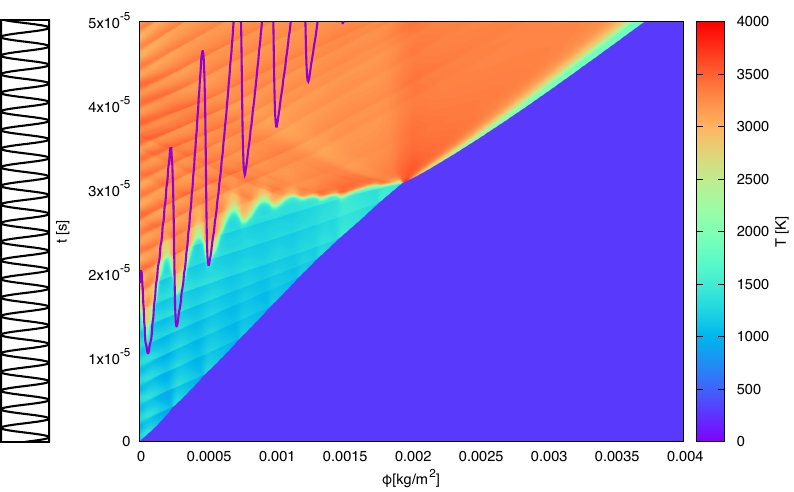}
\caption{Shock induced ignition in 2$H_2$+O$_2$ with $f=454$ kHz and $A=0.2 u_{p0}$; purple line denotes the ignition locus prediction without coupling to the gas exothermicity predicted by coupling of \eqref{eq:ignitionrate} and the analytical temperature field of section 4.}
\label{f542}
\end{center}
\end{figure}

\subsection{Forcing periods between $t_i$ and $t_r$}
\begin{figure}
\begin{center}
\includegraphics[width=0.7\textwidth]{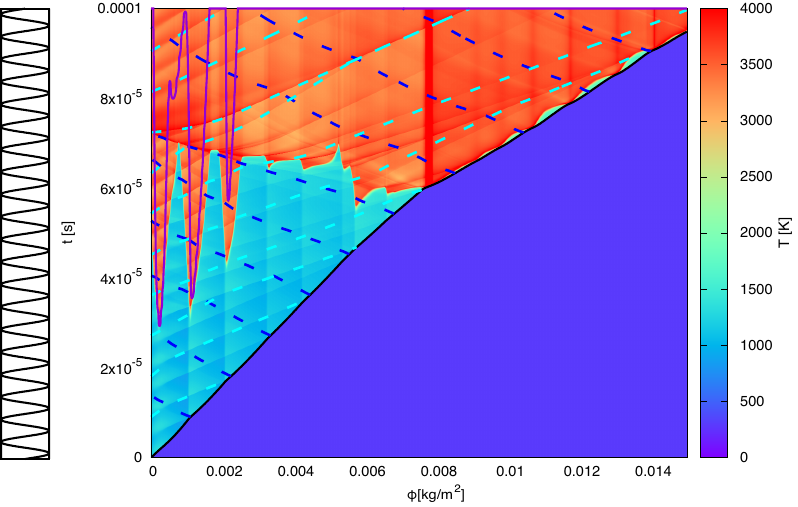}
\caption{Shock induced ignition in C$_2$H$_4$+3O$_2$ with $f=200$ kHz and $A=0.2 u_{p0}$; purple line denotes the ignition locus prediction without coupling to the gas exothermicity predicted by coupling of \eqref{eq:ignitionrate} and the analytical temperature field of section 4.}
\label{f56}
\end{center}
\end{figure}

When the forcing frequency is increased such that the nominal induction zone counts many forcing periods, a sequence of hot spots are observed. Figure \ref{f542} shows this interesting regime for the hydrogen-oxygen case for a frequency of $453.5kHz$ and amplitude $A=0.2 U_{p0}$. The higher frequency no longer significantly shortens the delay to first ignition at hot spots, since the inner shock amplitudes remains controlled by the fluctuation amplitude, which is kept constant.  The slight reduction in ignition delay in this case is due to the residual repeated shock energy dissipation modelled in the previous section. Instead, a series of sub-critical hotspots are generated along particle paths where internal shocks locally amplified the lead shock.  These hotspots can be clearly identified as vertical particle paths originating from the confluence of internal shocks with the lead shock.  Note that the higher frequency modulates the trajectory of the fast flame.  The saw-tooth fast flame trajectories are now significantly out of phase with the acoustics.  
This phenomenon is particularly striking in the more sensitive ethylene system illustrated in Fig.\ \ref{f56}. Owing to the much larger effective activation energy of the induction kinetics, the hotspots are igniting much earlier than cold spots and the sawtooth fast flames are even more out of phase with the acoustics. 

To help us in the interpretation of the mechanism of ignition, we made use of the temperature field prior to ignition determined in section 4, in order to determine the locus of the fast flame without account of exothermicity.  For this purpose, the induction layer is modelled in a conventional way by
\begin{equation}
\left(\frac{\partial \zeta}{\partial t}\right)_\phi=k \exp \left(-\frac{T_a}{T} \right) \label{eq:ignitionrate}
\end{equation}
where $\zeta$ is the induction progress variable ranging from 0 (fresh gases) to 1 (end of induction time) and the constants $k$ and $T_a$ are calibrated to recover the correct ignition delay and sensitivity to temperature.  These are calibrated for constant volume calculations in Cantera. They yield $T_a =9562.2$K and $k=1.6118\times 10^8 s^{-1}$ for the hydrogen-oxygen mixture and $T_a =23825.0$K and $k=3.452\times 10^{12} s^{-1}$ for the ethylene mixture at these operating conditions.  Given the temperature field, integration of \eqref{eq:ignitionrate} yields the ignition delay $t_i(\phi)$ along a particle path implicitly:
\begin{equation}
1=\int_{t_S}^{t_i}k \exp \left(-\frac{T_a}{T(t)} \right)\mathrm{dt}
\end{equation}
where $t_S(\phi)$ is the lead shock time. 

Figure \ref{f542} shows how the sequence of hotspots are formed and compares their location with the zero-order prediction.  While each sequential hotspot is formed from the coalescence of the internal shock with the lead shock, we note that the ignition delay of each hotspot is successively shorter than the previous one.  This is very well predicted by the ignition model with no exothermicity coupling.  The mechanism of this reduction is the cumulative energy dissipation heating along a particle path. Nevertheless, we observe that by the 4th hotspot, the energy release from previous hotspots has now played a sensible role in shortening the induction time.  The inert gasdynamic model begins to over-predict the ignition delay, since it neglects the effect of energy release on further gas compression in the induction zone.  The mechanism suggested from Fig.\ \ref{f542} is through the strengthening of forward facing shock waves passing through the main reaction zone. After the 6th hotspot, the sequence of hotspots become in phase with the motion of one of the inner shocks.  We thus see a discrete like gradient amplification mechanism where the sequence of hotspot onset becomes in phase with acoustics.  We label these \textit{hotspot cascades}. 

More complex hotspot cascades can be observed in more sensitive mixture. Figure \ref{f56} shows the ethylene-oxygen ignition with fluctuation frequency in $200kHz$ and amplitude in 20\% of the mean speed. The first ignition starts at $t=2.2\times 10^{-5}s$ form the first hotspot - this is very well predicted by the uncoupled ignition model. Nine hotspot ignitions can be observed prior to the reaction front merging to the lead shock and detonation formation. The second hot-spot is also very well predicted by the model.  The third, however, is accelerated by the energy release of the previous two, transmitted along the forward shocks originating from the first two. The three first hotspots thus follow the hotspot cascade mechanism described above.  Note that the inert model over-predicts the ignition delay more severely than for hydrogen, due to neglect of the gasdynamic heating from previous reactions.  By the fourth hot-spot, the heating provided by the previous three hot-spots gives rise to a more substantial decrease in the induction delay, as expected for this system with a higher activation energy. 

Nevertheless, the arrival of the first shock amplified by the hotspot cascade at the lead shock creates a strong enough disturbance to lead to prompt ignition at the hotspot along $\phi=0.006$ (hotspot 7).  This localized energy release triggers the backwards sequence of promoting the ignition of hotspot 6 and 5 and 4.  It also triggers the sequence of hotspots 8 and 9, aided by the shock from hotspot 3.  Note that the internal transitions follow again a discrete version of the gradient or SWACER mechanism.  We refer to this more complex situation as "bifurcated hot-spot cascade", since forward and backward hotspot avalanches are present due to hotspot - lead shock feedback.  

\subsection{Fast forcing}
\begin{figure}
\begin{center}
\includegraphics[width=0.7\textwidth]{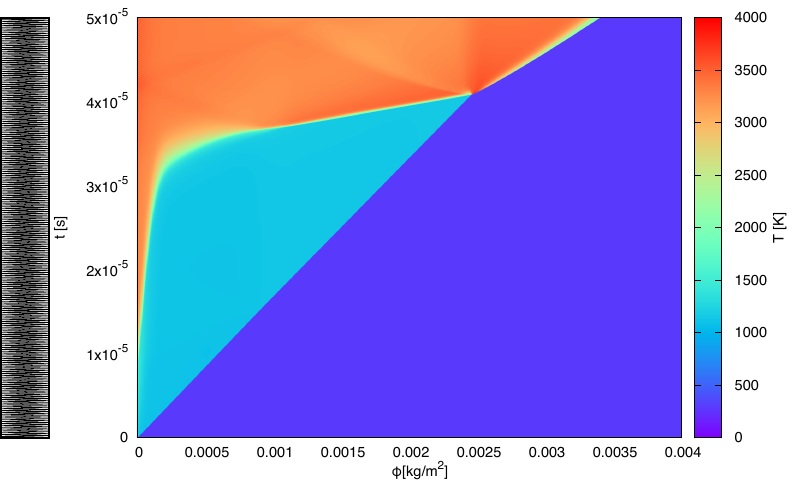}
\caption{Shock induced ignition in 2$H_2$+O$_2$ with $f=4540$ kHz and $A=0.2 u_{p0}$.}
\label{f57}
\end{center}
\end{figure}

Once the ignition delay is much longer than the fluctuation period, the hotspot cascade mechanism disappears. Figure \ref{f57} shows the pattern for hydrogen-oxygen ignition with frequency in $4535kHz$ and amplitude of 20\% of the mean speed. The first ignition starts near the piston at $t=1.0\times 10^{-5}s$ and a slow reaction front develops in the first layer of gas of width of approximately $\delta \phi \simeq 0.0002$ kg/m$^2$. Subsequently, the evolution follows a process very similar to the non-fluctuated case of Fig.\ \ref{f51.2}. The zone marked by the slow acceleration of the reaction front is likely due the finite region of dissipation created by the N-wave train. The N-wave train decays as $1/x$ and its influence is only felt a few wavelengths away from the piston.  For reference, the inert solution can be inferred from the results shown in Fig.\ \ref{f416} by rescaling the $t$ and $\phi$ by a factor of 10, controlled by the unique time scale in the inert problem, the fluctuation period. The inert N-wave train inferred by the rescaled results of Fig.\ \ref{f416} has decayed by nearly an order of magnitude at this penetration distance of $\delta \phi \simeq 0.0002$, which corresponds to 8 wavelengths of the train. The relatively slow reaction front observed in Fig.\ \ref{f57} is thus due to the energy dissipation gradient of the N-wave train decay.  Further away, the fluctuations no longer play a substantial role directly. Although the first ignition delay was very short, the detonation front forms relatively late. Compared to the non-fluctuation case (Fig.\ \ref{f51.2}), the inner supersonic reaction front forms almost at the same time. The place and time of merging between the inner supersonic reaction front and the lead shock does not have significant changes. This very high frequency fluctuation does not make a significant contribution to reducing the time of the detonation formation in this 1D ignition problem.\\
\begin{figure}
\begin{center}
\includegraphics[width=0.7\textwidth]{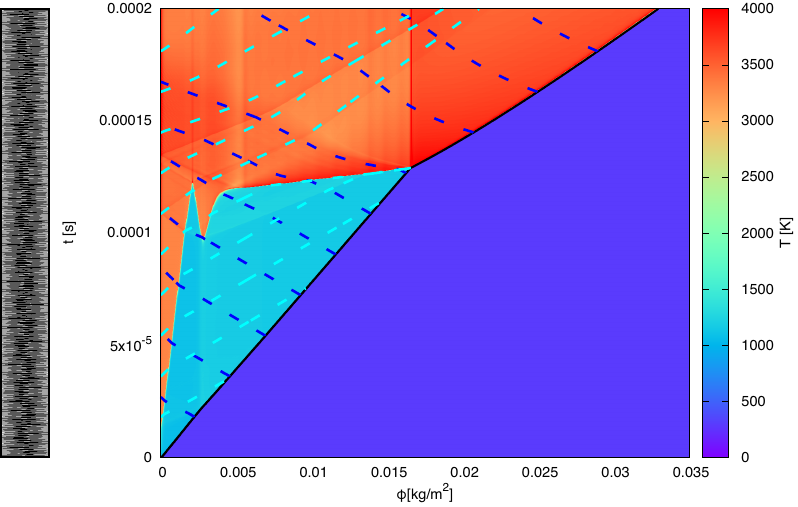}
\caption{Shock induced ignition in C$_2$H$_4$+3O$_2$ with $f=2000$ kHz and $A=0.2u_{p0}$.}
\label{f58}
\end{center}
\end{figure}

A similar pattern was observed for the more sensitive ethylene-oxygen system.  Figure \ref{f58} shows the ignition diagram with fluctuation frequency in $2000kHz$ and amplitude in 20\% of the mean speed. In this case, the ignition delay has been shortened at $t=1.3\times 10^{-5}s$ but the supersonic reaction wave takes longer to form. Two hotspots are formed in this case. The first hotspot is generated from the dissipation. The second hotspot formed by the complex hotspot cascade mechanism which is the interaction between the lead shock and intensified inner shock. In this mixture, the ignition delay is much more sensitive than the hydrogen-oxygen mixture. The intensified shock wave is strong enough to generate an ignition hotspot. This hotspot ignition locus quickly accelerate to an inner detonation, which catches up to the main shock.  These inner dynamics shorten the formation distance of the final detonation by approximately 90\% as compared to the nominal non-fluctuated case and cannot be neglected.

\section{Further discussion}

\begin{figure}
\begin{center}
\includegraphics[width=0.45\textwidth]{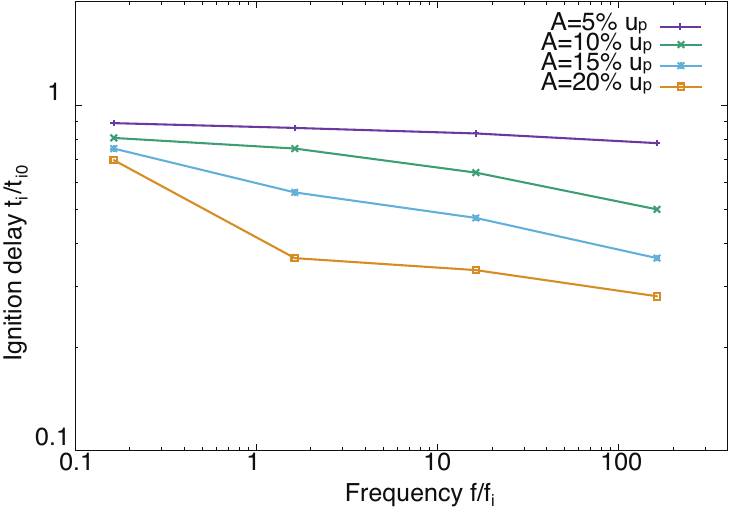}
\includegraphics[width=0.45\textwidth]{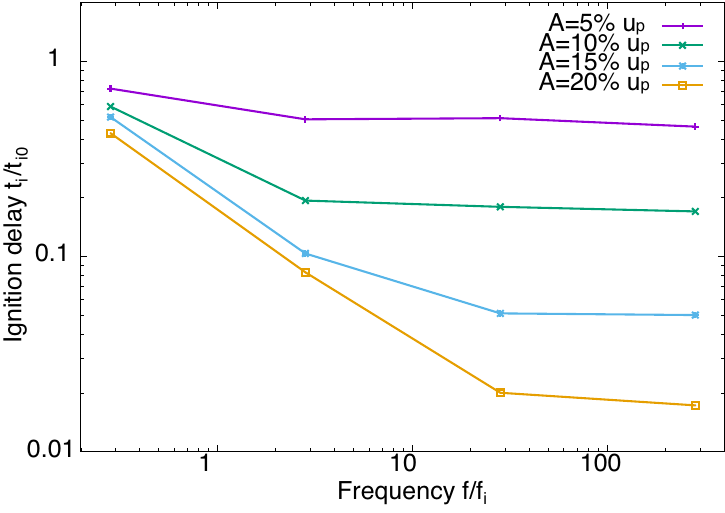}
\caption{Ignition delay of the first hot spot for 2$H_2$+O$_2$ (left) and C$_2$H$_4$+3O$_2$ (right) for different perturbation frequency and amplitude.}
\label{f61}
\end{center}
\end{figure}

The effect of the forcing frequency and forcing amplitude on the ignition delay of the first spot is shown in Fig.\ \ref{f61} for the two mixtures studied for 16 hydrogen-oxygen cases and 16 ethylene-oxygen cases. All the cases have been categorized into four groups by their forcing strength which ranges from 0.05$u_{p0}$ to 0.2$u_{p0}$. The ignition delay has been normalized by the original ignition delay without forcing: $t_{i0}$. The frequencies have been normalized by $f_i=1/t_{i0}$. These results were obtained numerically by solving the full Lagrangian problem.  These are well predicted by the ignition model without coupling, as discussed above. The effect of the forcing amplitude is to reduce the ignition delay, as it provides stronger hotspots, as expected.  The effect of the forcing frequency is however more interesting and reveals the role of the energy dissipation by shock heating.  It also shows that increasing frequency reduces the ignition delay time.  With increasing frequency, a particle of gas experiences more heating from the higher number of repeated shock interactions.  

The ignition delay is thus decreased by the increase in the cumulative energy dissipation.  Note also that the ignition delay reduction is much more substantial in the ethylene system, owing to the stronger sensitivity of ignition kinetics to temperature.   

The mechanisms of pressure wave enhancement identified in the present paper is in agreement with the gradient mechanism of Zel'dovich or the SWACER mechanism.  The physics operating in the low frequency regime is in line with the runaway mechanism for arbitrary gradients studied in the past by \cite{sharpe_short_2003}.  Changes in the forcing frequency changes the lead shock strength distribution, hence the induction delay distribution.  With increasing frequency, the fast flame from each distinct spot becomes less in phase with the acoustics.  The run-away process now relies on the sequence of hotspots being in phase with acoustics.  Two cascade mechanisms were identified. The first relies on the energy release of a previous hot-spot to strengthen a forward facing compression wave.  This in turn shortens the ignition delay of the next hot spot, and so-forth.  This mechanism is the discrete version of shock to detonation transition modelled by \cite{sharpe2002shock}.  

The second cascade mechanism occurs in sufficiently sensitive systems (high activation energy).  The first generation cascade amplifies a shock that arrives at the lead shock prior to the cascade culminating in transition to detonation.  The shock-shock interaction triggers a second generation of hotspots cascades of the first type.   

The phase of the perturbations induced by the piston was not studied in the present work.  When the forcing frequency is much larger than the inverse of ignition delay time, i.e., $f/f_i \gg 1$, the phase is not expected to play any sensible role.  However, when $f/f_i \lesssim 1$, the phase of the perturbation will likely play a very important one. For example, in the limit of $f/f_i \ll 1$, the problem consists of ignition and detonation transition behind a slowly accelerating or decelerating shock. When the shock accelerates, the negative temperature gradient behind it induces a more prompt ignition.  Likewise, ignition behind a decelerating shock is suppressed.  This type of problem can be best studied by the methods outlined by \cite{sharpe2002shock}. This is left for future study, since we are interested in the multi-hotspot scenarios leading to cooperative phenomena and cascades.  

\begin{figure}
\begin{center}
\includegraphics[width=0.45\textwidth]{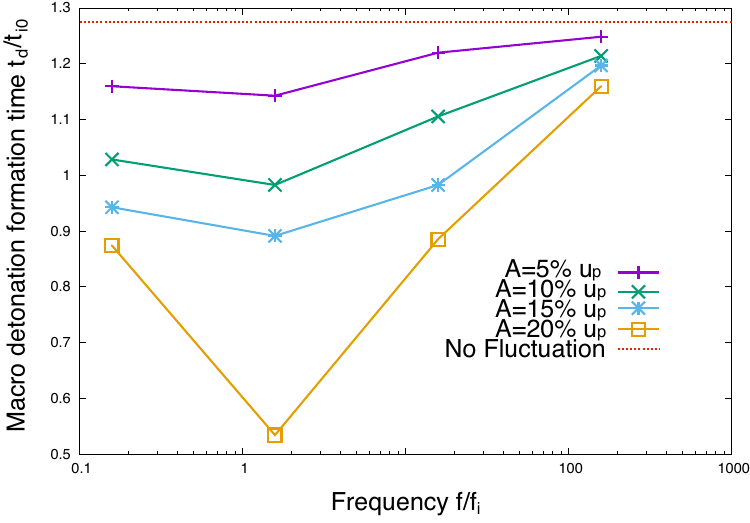}
\includegraphics[width=0.45\textwidth]{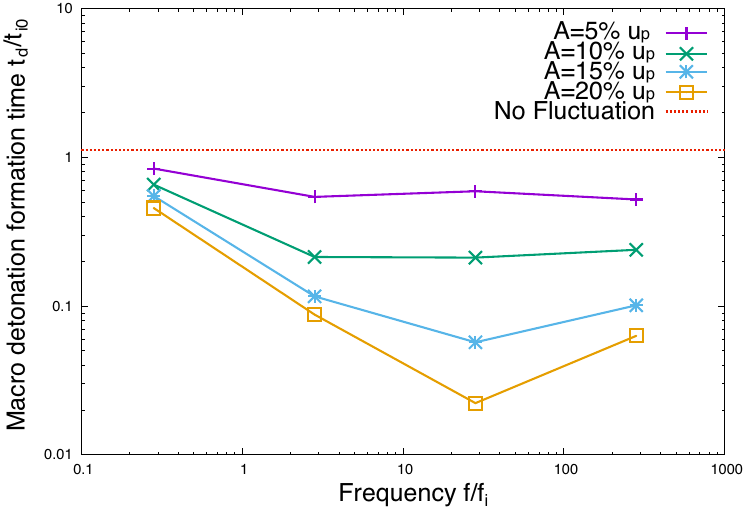}
\caption{Detonation formation time for 2$H_2$+O$_2$ (left) and C$_2$H$_4$+3O$_2$ (right) for different perturbation frequency and amplitude.}
\label{f64}
\end{center}
\end{figure}

Interestingly, the hotspot cascades in the multi-hot spot scenarios permit for detonations to form on the same time scales as the shortest ignition delay.  Figure \ref{f64} shows the detonation formation time for all the cases studied. These time scales are comparable with those of Fig.\ \ref{f61}.  The gasdynamic model developed is thus sufficient to determine this time of first ignition.   It is only for sufficiently high frequency oscillations that the cascades are suppressed by the mechanism of inner N-wave decay, reducing the effective penetration distance of the energy dissipation, as discussed above.  The inner shock decay results in a more pronounced gradient in hot-spot ignition gradient, suppressing the coherence between hot-spots and shocks.  At these high frequencies, the time to detonation actually increases with increasing frequency.  In the limit of high frequency, only one generation of hotspots near the piston remains, leading to prompt first ignition but unfavourable gradient.  

While the problem studied is a very canonical one, it may find direct application in the understanding of the late phases of deflagration to detonation transition, where transonic turbulent flames are generated \citep{saif2017chapman, poludnenko2019unified}.  A highly non-steady turbulent flame generating high frequency mechanical oscillations may trigger such hot-spot cascades.  When they are transonic, parts of their energy release is in phase with the acoustics.  Likewise, unstable detonations rely on high frequency oscillations to intermittently re-amplify and survive quenching \citep{radulescu2023evidenceP, sow2023dynamics}.  An analogous problem to our study is the problem of shock induced ignition in shock tubes by reflected shocks.  In this problem, boundary layers form on the walls, which provide the source of acoustic perturbations of transverse nature.  In this problem, premature ignition is often observed, and its origin is difficult to explain.  Perhaps the main advance in this problem is the study of \citet{khokhlov2015developmentP}, who first established that perturbations due to boundary layers do not directly control the onset of hot spots.  Instead, their continuous feedback to the main shock via acoustic waves provides the entropy perturbations favourable to the generation of hotspots. We mention this to highlight the long range effects of hotspots through interactions with the main shocks via acoustic waves, as discussed in the present study in the context of hotspot cascades. 

\section{Conclusion}
Our study of ignition behind a shock driven by an oscillating piston has revealed the intricate role of mechanical fluctuations on the ignition process and run-away to detonation.  A substantial decrease in the ignition delay was observed, owing to the changes in the lead shock strength.  With increasing frequency of oscillation, the role of lead shock modulation was decreased in favour of the role of the internal shock wave motion and resulting dissipative heating.  The internal shock wave motion permitted hotspot cascades, the organization of sequential hot spots in phase with the acoustics.  However, further increase in forcing frequency suppressed the cascade mechanism.  

The present paper shows conclusively that the high frequency forcing can play a very substantial role in the transition from a shock-induced ignition to detonation.  The results may thus clarify the general experimental observations of anomalous ignition in deflagration to detonation transition in turbulent flames and analogous phenomena within very unstable detonation waves.   

The present paper also provided significant advancements in the modelling of these events.  We provide a novel closed form model in Lagrangian coordinates for the generation of the train of compression and expansions, their steepening into a train of N-shock waves and their reflection on the lead shock, as well as the distribution energy dissipation rate in the induction zone.  The predictions of the model were found in excellent agreement with numerics.  The analytical approach permitted insight into the physical processes controlling the evolution of the temperature distribution in the induction zone prior to ignition. 

To the best of our knowledge, the formulation of the problem in Lagrangian coordinates, and the development and verification of the numerical scheme for reactive Lagrangian gas dynamics with multiple components and state dependent properties is also novel.

\section*{Declaration of Interests}
The authors report no conflict of interest.

\section*{Acknowledgements} 

We acknowledge financial support provided by the Natural Sciences and Engineering Research Council of Canada (NSERC) and Shell via a Collaborative Research and Development Grant, "Quantitative assessment and modelling of the propensity for fast flames and transition to detonation in methane, ethane, ethylene and propane”, with Dr.\ Andrzej Pekalski as research facilitator, and NSERC through the Discovery Grant "Predictability of detonation wave dynamics in gases: experiment and model development" (grant number RGPIN-2017-04322).  The last author also acknowledges financial support from AFOSR grant FA9550-23-1-0214 (program officer Dr.\ Chiping Li).

\bibliographystyle{jfm}
\bibliography{jfm-references_matei}
\appendix
\section{Transformation of the Euler equations for a multi-component reactive gas to Lagrangian coordinates}

We derive the Lagrangian form of the governing equations for a multi-component reactive medium and their characteristic form, generalizing the results obtained by \citet{fickett1971} for a single irreversible reaction.
  
\subsection{The reactive Euler equations in lab-coordinates}
The reactive Euler equations are well known in Eulerian, or lab coordinates.  In 1D, the independent variables are $x$, and $t$, time. The Euler equations can be written as: 
\begin{align}
 \frac{D \rho}{Dt} +\rho \left(\frac{\partial u}{\partial x}\right)_t=0\label{eq:eulercontinuity}\\
\rho \frac{D u}{Dt}+\left( \frac{\partial p}{\partial x} \right)_t=0 \label{eq:eulermomentum}\\
\rho \frac{D e_{tot}}{Dt}+\left(\frac{\partial \left(pu\right) }{\partial x}\right)_t=0 \label{eq:eulerenergy}\\
\rho \frac{D Y_i}{Dt}= \omega_i \label{eq:eulerspecie}
\end{align}
The material derivative is $\frac{D}{Dt} = \left(\frac{\partial }{\partial t}\right)_x+u \left(\frac{\partial }{\partial x}\right)_t$, $e_{tot}=e+\frac{1}{2}u^2$ is the total energy, $e$ the specific internal energy of the mixture, $Y_i$ the mass fraction of the i$^\text{th}$ component and $\omega_i$ the mass production rate of specie  $i$ per unit volume, per unit time, obtained from chemical kinetics.   After some thermodynamic manipulations, the energy equation \eqref{eq:eulerenergy} can be also written as 
\begin{equation}
\frac{D p}{Dt}=c^2\frac{D \rho}{Dt}+\rho c^2 \dot{\sigma} \label{eq:eulerenergythermicity}\\
\end{equation}
where $\dot{\sigma}$ is the thermicity, which denotes the gasdynamic effect of chemical reactions or other relaxation phenomena on the rate of pressure and speed changes along the family of characteristics.  The thermicity in its most general form is given by:
\begin{equation}
\dot{\sigma}= - \frac{\rho}{c_p} \left(\frac{\partial v}{\partial T}\right)_{p, Y_i} \sum_{i=1}^N \left( \frac{\partial h}{\partial Y_i}\right)_{p, \rho, Y_{j,j \neq i}}\frac{D Y_i}{D t}
  \label{eq:thermicity}
\end{equation}
\citep{Fickett&Davis1979}.  For a mixture of ideal gases, the thermicity reduces to
\begin{equation}
\dot{\sigma}= \sum\limits_{i=1}^{N}\left(\frac{\bar{W}}{W_i}-\frac{h_i}{c_pT}\right) \frac{D Y_i}{D t}
\end{equation}
where $W_i$ is the molecular weight of the i$^\text{th}$ component, $\bar{W}$ is the mean molecular weight of the mixture, $h_i$ is the specific enthalpy of the i$^\text{th}$ specie and $c_p$ is the mixture frozen specific heat. 

The characteristic equations are obtained by simple manipulations of these expressions and yield
\begin{equation}
\frac{D_\pm p}{Dt} \pm \rho c \frac{D_\pm u}{Dt}=\rho c^2 \dot{\sigma} \label{eq:eulercharacteristics}\\
\end{equation}
where 
\begin{equation}
\frac{D_\pm }{D t} \equiv \left( \frac{\partial }{\partial t} \right)_x +( u\pm c) \left( \frac{\partial }{\partial x}\right)_t 
\end{equation}
 are derivatives along the C+ and C- characteristics, given respectively by $dx/dt =u\pm c$. 
\subsection{Lagrangian coordinates and transformation rules} 
In 1D problems where one boundary condition can be prescribed along a particle path $x_p(t)$, where $u=\dot{x}_p(t)$, $\rho=\rho_p(t)$, etc., we can change from an Eulerian frame to a Lagrangian frame, by transforming the reactive Euler equations expressed with independent variables $(x,t)$ to Lagrangian independent variables $(\phi, t')$ by the formal change of variables:
\begin{equation}
\phi=\int_{x_p(t)}^x\rho \mathrm{d}x, \qquad t'=t \label{eq:lagrangiantransformation}
\end{equation}
The density weighted coordinate $\phi$ remains constant along a particle path through the conservation of mass, as we will show below.  Equations \eqref{eq:lagrangiantransformation} permit to evaluate the derivatives $\left(\partial \phi / \partial x\right)_t$ and $\left(\partial \phi / \partial t\right)_x$ required for the change of variables.  Using the Leibniz rule of differentiation of an integral where both the integrand and the integral bounds vary with the variable used for differentiation, we obtain:
\begin{equation}
\left(  \frac{\partial \phi}{\partial x} \right)_t = \rho, \qquad \left(   \frac{\partial \phi}{ \partial t} \right)_x=\int_{x_p(t)}^x \left( \frac{\partial\rho}{\partial t} \right)_x \mathrm{d}x-\rho_p \dot{x}_p \label{eq:lagrangiantransformation2}
\end{equation}
To evaluate the last integral in \eqref{eq:lagrangiantransformation2}, we make use of the continuity equation \eqref{eq:eulercontinuity} re-written as,
\begin{equation}
\left( \frac{\partial \rho}{\partial t} \right)_x + \left( \frac{\partial \rho u}{\partial x} \right)_t = 0 \label{eq:eulercontinuity2}
\end{equation}
and obtain
\begin{equation}
\left(  \frac{\partial \phi}{\partial x} \right)_t = \rho, \qquad \left(   \frac{\partial \phi}{ \partial t} \right)_x=-\rho u  \label{eq:lagrangiantransformation3}
\end{equation}
Using \eqref{eq:lagrangiantransformation3}, it can be verified that the variation of $\phi$ along a particle path, namely, 
\begin{equation}
\frac{D \phi}{D t}\equiv\left( \frac{\partial \phi}{\partial t}\right)_x + u \left( \frac{\partial \phi}{\partial x}\right)_t = 0
\end{equation}
is indeed zero and $\phi$ serves as a particle label.

We can now operate the formal change of variables using \eqref{eq:lagrangiantransformation3} and the chain rule of differentiation for any field variable $a \left(\phi(x, t), t'(x,t) \right) $, yielding the following:
\begin{equation}
\left(\frac{\partial a}{\partial x} \right)_t=\left(\frac{\partial a}{\partial \phi} \right)_{t'} \left(\frac{\partial \phi}{\partial x} \right)_t +  \left(\frac{\partial a}{\partial t'} \right)_\phi \left(\frac{\partial t'}{\partial x} \right)_t=\rho \left(\frac{\partial a}{\partial \phi} \right)_{t'} \label{eq:EtoL1}
\end{equation}
\begin{equation}
\left(\frac{\partial a}{\partial t} \right)_x=\left(\frac{\partial a}{\partial \phi} \right)_{t'} \left(\frac{\partial \phi}{\partial t} \right)_x +  \left(\frac{\partial a}{\partial t'} \right)_\phi \left(\frac{\partial t'}{\partial t} \right)_x=-\rho u \left(\frac{\partial a}{\partial \phi} \right)_{t'} + \left(\frac{\partial a}{\partial t'} \right)_{\phi}\label{eq:EtoL2}
\end{equation}
The above two expressions can also be used for substitution expressions for derivatives along particle paths: 
\begin{equation}
\frac{D a}{D t}\equiv\left( \frac{\partial a}{\partial t}\right)_x + u \left( \frac{\partial a}{\partial x}\right)_t = \left(\frac{\partial a}{\partial t'} \right)_{\phi}\label{eq:EtoL3}
\end{equation}
and along C+ and C- characteristics:
\begin{equation}
\frac{D_\pm a}{D t} \equiv \left( \frac{\partial a}{\partial t}\right)_x +( u\pm c) \left( \frac{\partial a}{\partial x}\right)_t = \left(\frac{\partial a}{\partial t'} \right)_{\phi} \pm \rho c \left(\frac{\partial a}{\partial \phi} \right)_{t'}\label{eq:EtoL4}
\end{equation}

\subsection{The reactive Euler equations in Lagrangian coordinates}

Equations \eqref{eq:EtoL1} to \eqref{eq:EtoL4} now permit to re-write the reactive Euler equations \eqref{eq:eulercontinuity}-\eqref{eq:eulerspecie} as:
\begin{align}
\left( \frac{\partial v}{\partial t'} \right)_\phi- \left( \frac{\partial u}{\partial \phi} \right)_{t'}=0 \label{eq:lagrangiancontinity}\\
\left( \frac{\partial u}{\partial t'} \right)_\phi + \left( \frac{\partial p}{\partial \phi} \right)_{t'}=0 \label{eq:lagrangianmomentum}\\
 \left( \frac{\partial e_{tot}}{\partial t'} \right)_\phi + \left( \frac{\partial pu }{\partial \phi} \right)_{t'}=0 \label{eq:lagrangianenergy}\\
\left( \frac{\partial Y_i}{\partial t'} \right)_\phi= \frac{\omega_i}{\rho} \label{eq:lagrangianspecie}
\end{align}
where $v=1/\rho$ is the specific volume.

The alternate form of the energy equation \eqref{eq:eulerenergythermicity} becomes
\begin{equation}
\left( \frac{\partial p}{\partial t'} \right)_\phi=c^2 \left( \frac{\partial \rho}{\partial t'} \right)_\phi+\rho c^2 \dot{\sigma} \label{eq:lagrangianenergythermicity}\\
\end{equation}
and the characteristic equation \eqref{eq:eulercharacteristics} becomes
\begin{equation}
\left( \left(\frac{\partial }{\partial t'}\right)_\phi \pm \rho c \left(\frac{\partial }{\partial \phi}\right)_{t'} \right)p \pm \rho c \left( \left(\frac{\partial }{\partial t'}\right)_\phi \pm \rho c \left(\frac{\partial }{\partial \phi}\right)_{t'} \right)u=\rho c^2 \dot{\sigma} \label{eq:lagrangiancharacteristics}\\
\end{equation}
We remark that the C+ and C- characteristics are given by  
\begin{equation}
\frac{\mathrm{d} \phi}{\mathrm{d} t'}=\pm \rho c
\end{equation}
and the characteristic speeds are $\pm \rho c$.  

\section{Reflected disturbances at the lead shock}
\label{app:shockcatchup}
\begin{figure}
\begin{center}
\includegraphics[width=0.5\textwidth]{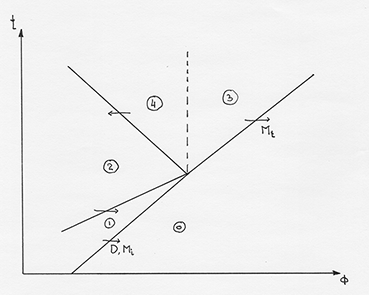}
\caption{A disturbance catching up to the lead shock, modifying its strength and resulting acoustic and entropy wave generation.}
\label{fig:shockcatchup}
\end{center}
\end{figure}

We consider the problem of a weak disturbance propagating along a C+ characteristic overtaking a lead shock of arbitrary strength, generating a reflected acoustic disturbance along a C- characteristic and a entropy wave along a particle path. Fig.\ \ref{fig:shockcatchup} illustrates the various states:  State 1 is the post shock state, state 2 is the post incident disturbance state, state 4 is the state behind the reflected acoustic disturbance and state 3 is the state behind the modified lead shock.  The upstream uniform state 0 and states 1 and 2 are known and we seek to determine states 3 and 4.  The original incident shock has Mach number $M_i$, and the new Mach number of the incident shock is disturbance is $M_t$.  We characterize the inner disturbance by its overpressure, which we take as being a small perturbation $z_{12}=(p_2-p_1)/p_1\ll1$.  We thus linearize around state 1, i.e.,
\begin{align}
p=p_1+ p'\\
\rho = \rho_1 + \rho'\\
T = T_1 + T'\\
M_t=M_0+ M'
\end{align}
where the prime quantities are small.

The linearised version of the characteristic equations along the C$\pm$ characteristics becomes:
\begin{align}
d p' \pm \rho_1 c_1 d u' = 0 \label{eq:linearCcompatibility}
\end{align}
The energy equation for a non-reacting perfect gas along a particle path is
\begin{align}
\frac{dp}{p}-\frac{\gamma}{\gamma-1}\frac{dT}{T}=0
\end{align}
which linearises to: 
\begin{align}
\frac{p'}{p_1} -\frac{\gamma}{\gamma-1}\frac{T'}{T_1}=0
\end{align}
The disturbances in state 3 are related to the change in Mach number via the Rankine-Hugoniot relations:
\begin{align}
p_3' = \left(\frac{dp}{dM}\right)_{RH} M'\\
u_3' = \left(\frac{du}{dM}\right)_{RH} M'\\
T_3' = \left(\frac{dT}{dM}\right)_{RH} M'
\end{align}
where derivatives with subscript RH are to be taken from the Rankine-Hugoniot relations $p(M)$, $u(M)$ and $T(M)$ respectively.  Below, these derivatives are taken with respect to pressure, e.g., $(du/dM)_{RH}=(du/dp)_{RH} (dp/dM)_{RH}$.

Using the method of characteristics, the pressure, speed and temperature disturbances can be found by straightforward solution of the compatibility equations \eqref{eq:linearCcompatibility} in the acoustic regime considered.  Across the incident disturbance, the C- compatibility relation requires 
\begin{align}
p_1'- \rho_1 c_1 u_1' = 0 
\end{align}
Across the reflected disturbance, the C+ compatibility relation requires
\begin{align}
p_2'- \rho_1 c_1 u_2' =  p_4'- \rho_1 c_1 u_4'
\end{align}
across the contact discontinuity, mechanical equilibrium applies, $u'_3=u'_4$ and $p'_3=p'_4$.  
Solving this system of algebraic equations, one obtains the strength of the disturbances in terms of $p_2'$:
\begin{align}
u_2'=\left(\frac{1}{\rho_1 c_1} \right) p_2' \\
T_2'=\left(\frac{\gamma-1}{\gamma}\frac{T_1}{p_1} \right) p_2' \\
p_3'=p_4'=\left(\frac{2}{1+\rho_1 c_1 (du/dp)_{RH}} \right) p_2'\\
u_3'=u_4'=\left(\frac{2 (du/dp)_{RH}}{1+\rho_1 c_1 (du/dp)_{RH}} \right) p_2'\\
T_3'=\left(\frac{2 (dT/dp)_{RH}}{1+\rho_1 c_1 (du/dp)_{RH}} \right) p_2'\\
T_4'=\left(\frac{\gamma-1}{\gamma}\frac{T_1}{p_1}\frac{2}{1+\rho_1 c_1 (du/dp)_{RH}}  \right) p_2' 
\end{align}

\begin{figure}
\begin{center}
\includegraphics[width=0.7\textwidth]{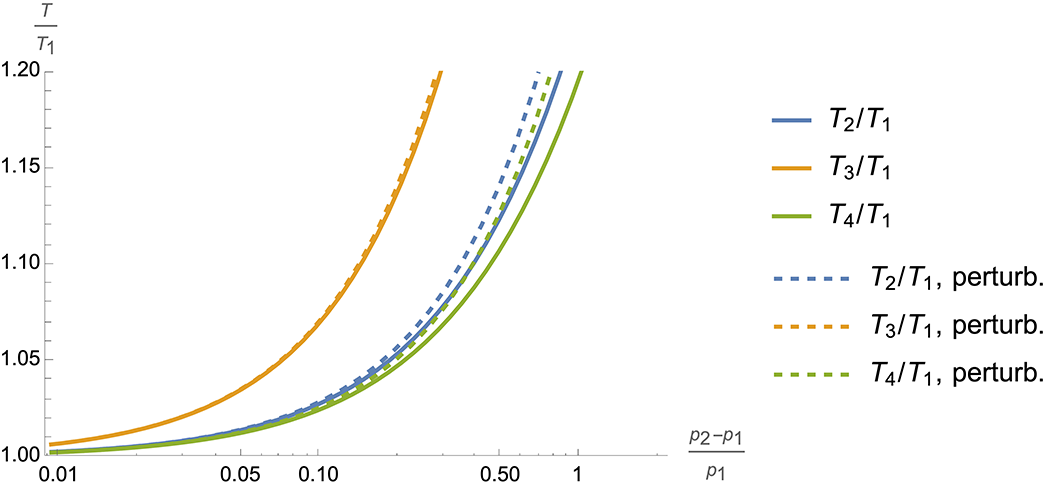}
\caption{Temperature perturbations behind a $M_i=4.05$ lead shock after the catch up of an arbitrary strength shock in a perfect gas with $\gamma=1.348$; full lines is the exact solution while broken lines is the acoustic approximation for the inner disturbance.}
\label{fig:shockcatchup2}
\end{center}
\end{figure}

This completes the solution.  Figure \ref{fig:shockcatchup2} shows the performance of the acoustic treatment of inner perturbations on the temperatures obtained behind the shock.  Comparison is made against exact calculations using the exact jump equations where the inner disturbance is a shock wave and the reflected wave is a centred expansion wave \citep{bull1953interaction}.
  
Simplifications for weak or strong incident shocks are straightforward.  If one assumes a strong lead shock, starting with the usual strong shock RH expressions
\begin{align}
\frac{u}{D}=\frac{p}{\rho_0 D^2}=\frac{2}{\gamma+1}  \quad \textrm{and} \quad  \frac{\rho}{\rho_0}=\frac{\gamma+1}{\gamma-1}
\end{align}
one obtains 
\begin{align}
p_3'=p_4'&= \frac{4(\gamma-1)}{2(\gamma-1)+\sqrt{2\gamma(\gamma-1)}}  p_2'\\
u_3'=u_4'&=  \frac{1}{1+\frac{\gamma}{\sqrt{2\gamma(\gamma-1)}}}  \frac{p_2'}{\rho_0 D}\\
\frac{T_2'}{T_1}&=\frac{\gamma-1}{\gamma} \frac{p_2'}{p_1}\\
\frac{T_3'}{T_1} &= \frac{4 (\gamma-1)}{2(\gamma-1)+\sqrt{2\gamma(\gamma-1)}}  \frac{p_2'}{p_1}\\
\frac{T_4'}{T_1}&=  \frac{4(\gamma-1)^2}{\gamma \left(2(\gamma-1)+\sqrt{2\gamma(\gamma-1)}\right)}   \frac{p_2'}{p_1} 
\end{align}

For weak shocks treated in this paper, the exact Rankine-Hugoniot shock jump conditions can be parametrized by the shock overpressure $z=(p-p_0)/p_0$ and truncated at the desired order.  The exact expressions are given by \citet{whithambook} :
\begin{align*}
\frac{p}{p_0}=1+z, \quad
\frac{\rho}{\rho_0}=\frac{1+\frac{\gamma+1}{2\gamma}}{1+\frac{\gamma-1}{2\gamma}}, \quad
\frac{c}{c_0}=\sqrt{\frac{p}{p_0}\frac{\rho_0}{\rho}}, \quad
\frac{T}{T_0}=\left( \frac{c}{c_0} \right)^2
\end{align*} 
\begin{align}
\frac{u-u_0}{c_0}=\frac{z}{\gamma \sqrt{1+\frac{\gamma+1}{2\gamma}}}, \quad
M=\sqrt{1+\frac{\gamma+1}{2\gamma}}
\end{align}
which permits to re-write the derivatives in terms of $z$, for example
\begin{align}
\left(\frac{du}{dp}\right)_{RH}=\frac{\left(\frac{du}{dz}\right)_{RH}}{\left(\frac{dp}{dz}\right)_{RH}}=\frac{c_0}{p_0}\frac{\sqrt{2+z+z/\gamma}\left(z+\gamma(4+z)\right)}{\sqrt{2}\left(z+\gamma(2+z)  \right)^2}
\end{align}
If one wishes to truncate the approximation to weak incident shocks, while retaining non-linearity, the RH equations can be expanded in Taylor series in terms of $z$ and retain only terms up to $O(z^2)$, such that entropy and the Riemann variable $J^-$ remain constant; the leading contribution to changes in entropy and $J^-$ across the shock are terms of order $O(z^3)$ \citep{whithambook}. 

\section{Grid convergence study and the effect of numerical resolution}
\label{app:convergence} 
A grid convergence study was performed for the numerically challenging problems of ignition with embedded hotspots leading to hotspot cascades.  The conditions targeted are those illustrated in Figs.\ \ref{f542} and \ref{f56} for the hydrogen and ethylene mixtures respectively.

\begin{figure}
\begin{center}
\includegraphics[width=1\textwidth]{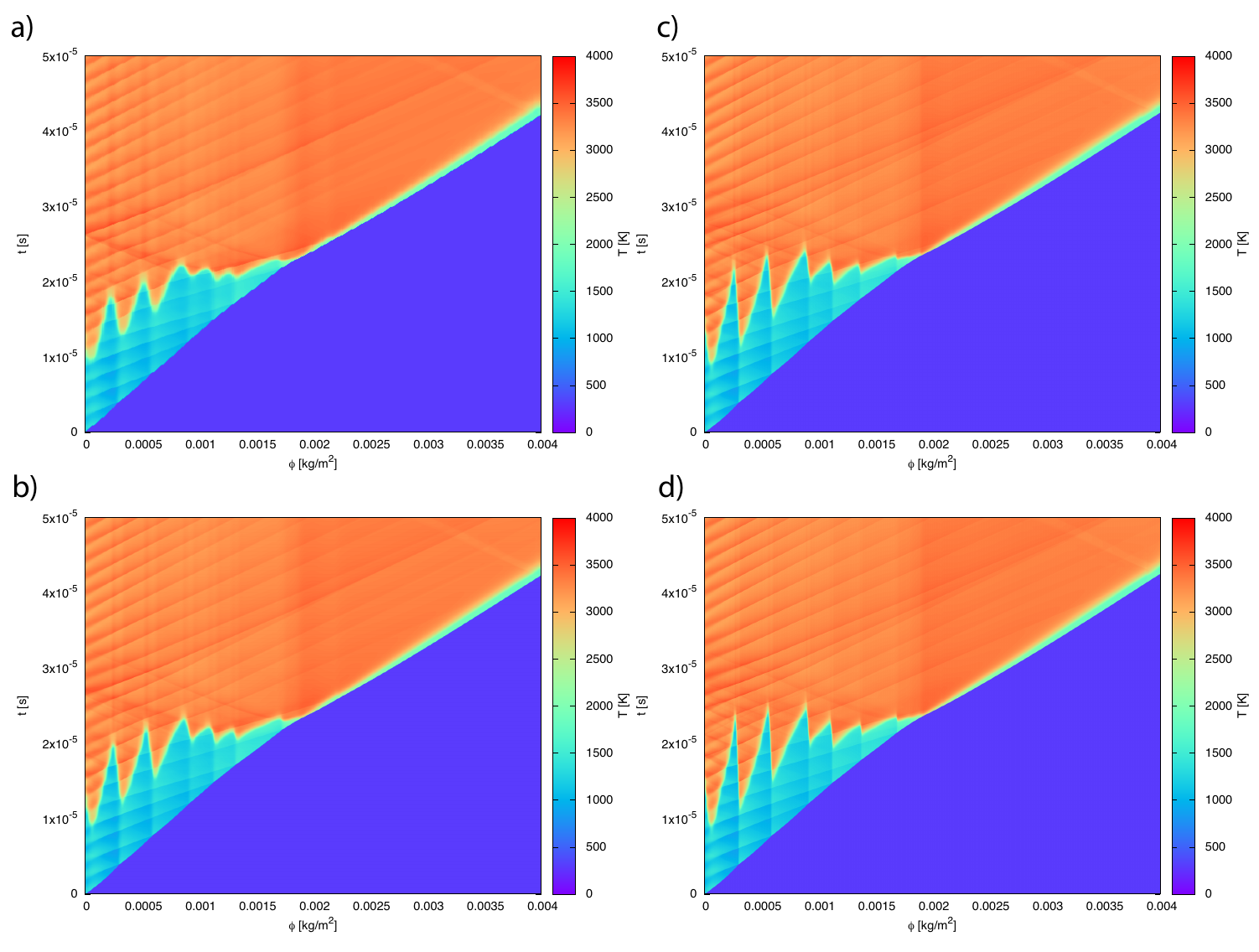}
\caption{Grid convergence study for ignition in 2$H_2$+O$_2$ with $f=454$ kHz and $A=0.2 u_{p0}$ for a grid spacing of $\Delta_{\phi}$ of a) $8\times10^{-6}$kg/m$^2$, b) $4\times10^{-6}$kg/m$^2$, c) $2\times10^{-6}$kg/m$^2$ and d) $1\times10^{-6}$kg/m$^2$. }
\label{fig:convergenceH2}
\end{center}
\end{figure}

Figure \ref{fig:convergenceH2} shows the numerical results obtained for 500, 1000, 2000 and 4000 grid points covering the $\phi$ axis, corresponding respectively to a grid spacing of $\Delta_{\phi}$ of $8\times10^{-6}$kg/m$^2$, $4\times10^{-6}$kg/m$^2$, $2\times10^{-6}$kg/m$^2$ and $1\times10^{-6}$kg/m$^2$.  As can be verified, the effect of changing the resolution over an order of magnitude has minimal impact on the solution, both quantitatively and qualitatively.  The principal difference is the reduction of numerical dissipation at contact surfaces resulting from the shocks overtaking the lead shock.  While the rear facing (non-physical) flames are clearly observed at the lowest resolution along these contact surfaces, these are significantly minimized as the resolution is increased and the physical fast flames propagating forward have speeds an order of magnitude larger than numerical ones. At the largest resolutions of case c) and d), the reaction profile acquires its characteristic saw-tooth profile with nearly vertical contact surfaces.  The resolution reported in the results of the paper use the resolution as in c).  

It is worthwhile commenting on the effect of resolving the inner structure of the internal fast flames.  Their characteristic thickness is given by the product of the wave speed and the characteristic reaction time $t_r$ listed in Table 1.  Taking the shock speed as the characteristic speed, for reference, the resulting characteristic thickness in this case is $\delta \phi_{r}\simeq 1\times10^{-4}$kg/m$^2$.  This means that the solutions presented above have 13 points per $\delta \phi_{r}$ at the lowest resolution and up to 100 points per $\delta \phi_{r}$ at the highest resolution.  As internal waves propagate at lower speeds, the number of points in their thickness degrades in the same proportion as the speed drops.

\begin{figure}
\begin{center}
\includegraphics[width=1\textwidth]{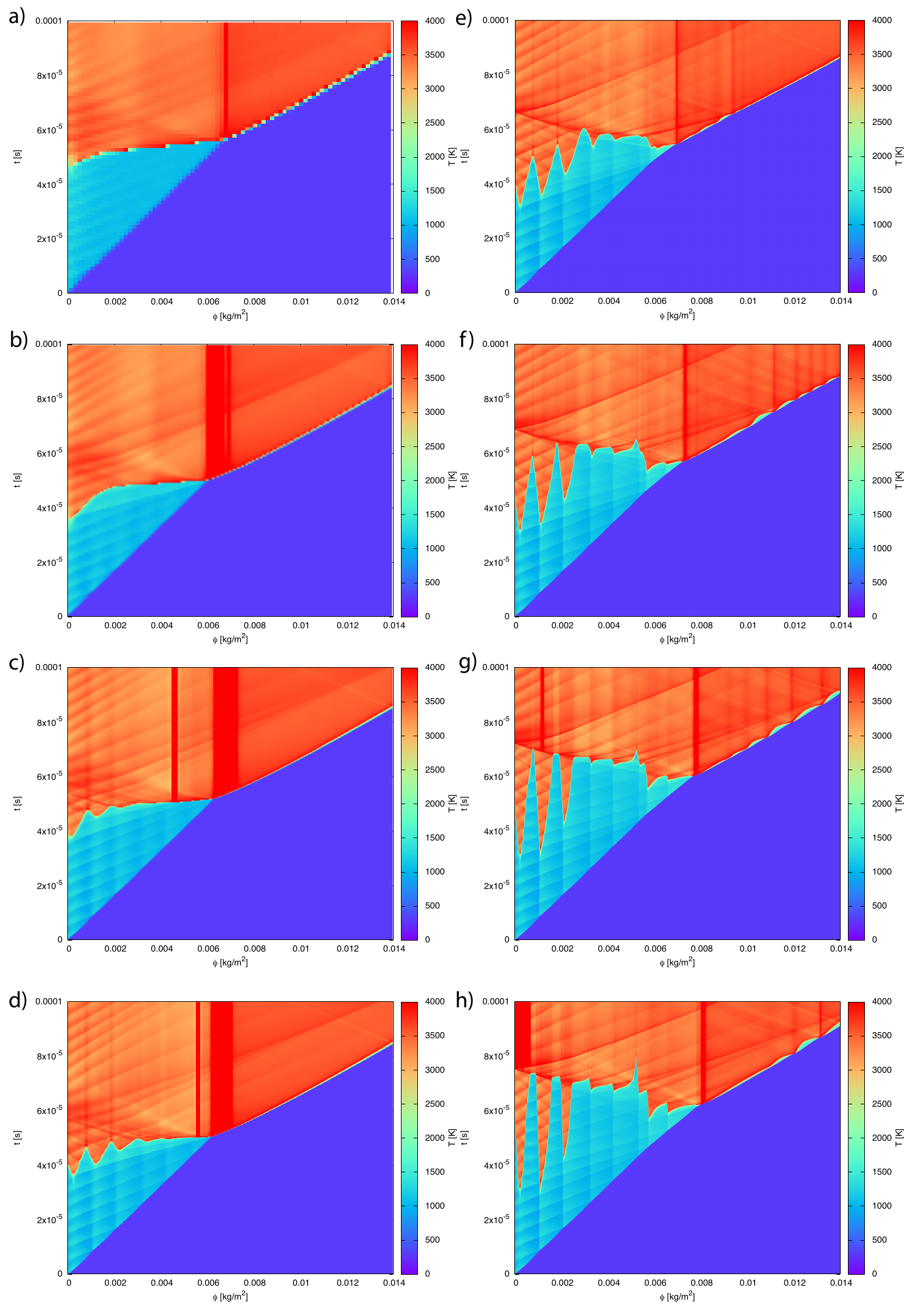}
\caption{
Grid convergence study for ignition in C$_2$H$_4$+3O$_2$ with $f=200$ kHz and $A=0.2 u_{p0}$ for a grid spacing of $\Delta_{\phi}$ of a) $1.8\times10^{-4}$kg/m$^2$, b) $9.0\times10^{-5}$kg/m$^2$, c) $4.5\times10^{-5}$kg/m$^2$, d) $2.2\times10^{-5}$kg/m$^2$ e) $1.1\times10^{-5}$kg/m$^2$, f) $5.6\times10^{-6}$kg/m$^2$, g) $2.8\times10^{-6}$kg/m$^2$ and h) $1.4\times10^{-6}$kg/m$^2$. }
\label{fig:convergenceC2H4}
\end{center}
\end{figure}

The resolution study performed for ignition in ethylene is more revealing of the importance of resolving the very thin reaction zones.  Figure \ref{fig:convergenceC2H4} shows the numerical results obtained for varying the resolution over two orders of magnitude, frame g) corresponding to the resolution of the results presented in the paper.  The number of points in the domain were respectively 78, 156, 312, 625, 1250, 2500, 5000 and 10000 grid points covering the $\phi$ axis, corresponding respectively to grid spacings $\Delta_{\phi}$ of $\Delta_{\phi}$ of $1.8\times10^{-4}$kg/m$^2$, $9.0\times10^{-5}$kg/m$^2$, $4.5\times10^{-5}$kg/m$^2$, $2.2\times10^{-5}$kg/m$^2$, $1.1\times10^{-5}$kg/m$^2$, $5.6\times10^{-6}$kg/m$^2$, $2.8\times10^{-6}$kg/m$^2$ and $1.4\times10^{-6}$kg/m$^2$.  For reference, the characteristic reaction zone thickness is $\delta \phi_{r}\simeq 4.1\times10^{-5}$kg/m$^2$. This means that the most resolved simulation has approximately 30 points per $\delta \phi_{r}$ and cases a), b) and c) do not resolve the reaction zone thickness at all.  

Similar to the hydrogen case, the effect of increasing the resolution is to sharpen the contact surfaces and minimize the non-physical numerical flames propagating backwards.  At the low resolution of cases a) and b), the hotspots are not resolved at all, as numerical flames consume them as soon as they are present.  With increasing resolution, hotspots appear and burn progressively slower as the resolution is increased.  By cases g) and h), the fast flame structure acquires the expected saw tooth shape with minor differences in the sequence of events observed.  Note however that the lowest speed fast flames originating from the first three hotspots are still somewhat affected by numerical resolution.  Nevertheless, the solution remains qualitatively similar and described by the same physics discussed in the main text.  

The general conclusion of our resolution study suggest that at least approximately 10 grid points are required to cover the characteristic reaction zone thickness of fast flames in order to guarantee that the fast flames propagate well in excess of numerical flames.  This resolution was ensured in the body of this work.  In practice, lower resolutions lead to hotspots developing more rapidly due to numerical diffusion.  This is now well understood in the literature of numerical simulations of inviscid gasdynamic problems coupled to stiff reactive problems leading to contact surfaces and shear layers along which numerical flames are established \citep{GAMEZO1999154, sharpe2001transverse, radulescu2007hydrodynamic, radulescu2018detonation}.  

The discussion highlights the importance of resolving the reaction zone of fast flames, not only the induction zones, which is usually the metric used in the literature. 

\end{document}